\documentclass{aa}    
\usepackage{txfonts}
\usepackage{graphicx}
\usepackage{natbib}
\bibpunct[, ]{(}{)}{,}{a}{}{,}

\usepackage{color}

\newcommand{\logg}{\rm log~ g}
\newcommand{\Teff}{T_{\rm eff}}
\newcommand{\eps}[1]{\log\varepsilon_{\rm #1}}
\newcommand{\jj}[3]{\mbox{$(#1/2,#2/2)_#3$}}
\newcommand{\Eexc}{$E_{\rm exc}$}
\newcommand{\eu}[5]{\mbox{$#1\,^#2{\rm #3}^{#4}_{\rm #5}$}}
\newcommand{\kH}{$S_{\!\rm H}$}    
\newcommand{\kms}{km\,s$^{-1}$}
\newcommand{\Vmic}{V_{\rm mic}}
\newcommand{\Vmac}{V_{\rm mac}}

\begin{document}

\title{Non-LTE effects on the lead and thorium abundance determinations for cool stars}

\author{
L. Mashonkina\inst{1,2} \and  
A. Ryabtsev\inst{3} \and  
A. Frebel\inst{4,5} 
 }

\offprints{L. Mashonkina; \email{lima@inasan.ru}}
\institute{
     Universit\"ats-Sternwarte M\"unchen, Scheinerstr. 1, D-81679 M\"unchen, 
     Germany \\ \email{lyuda@usm.lmu.de}
\and Institute of Astronomy, Russian Academy of Sciences, RU-119017 Moscow, 
     Russia \\ \email{lima@inasan.ru}
\and Institute of Spectroscopy, Russian Academy of Sciences, 142190, Troitsk, Moscow region, Russia 
\and Massachussetts Institute of Technology,
Kavli Institute for Astrophysics and Space Research, 77 Massachusetts Avenue,
Cambridge, MA 02139, USA \\ \email{afrebel@mit.edu} 
\and Harvard-Smithsonian Center for Astrophysics, 60
Garden St, Cambridge, MA 02138, USA
}

\date{Received  / Accepted }

\abstract { Knowing accurate lead abundances of metal-poor stars
  provides constraints on the Pb production mechanisms in the early
  Galaxy. Accurately deriving thorium abundances permits a
  nucleo-chronometric age determination of the star.}  
{We aim to improve the calculation of the \ion{Pb}{i} and \ion{Th}{ii}
  lines in stellar atmospheres based on non-local thermodynamic
  equilibrium (non-LTE) line formation, and to evaluate the influence
  of departures from LTE on Pb and Th abundance determinations through
  a range of stellar parameters with variations of the metallicity
  from the solar value down to [Fe/H] = $-3$.}
{Comprehensive model atoms for \ion{Pb}{i} and \ion{Th}{ii} are
  presented. We describe calculations of the \ion{Pb}{i} energy levels
  and oscillator strengths.}
{The main non-LTE mechanism for \ion{Pb}{i} is the ultraviolet
  overionization. Non-LTE leads to systematically depleted total
  absorption in the \ion{Pb}{i} lines and accordingly, positive
  abundance corrections. The departures from LTE grow with decreasing
  metallicity. Non-LTE removes the discrepancy between the solar
  photosphere and the meteoritic lead abundance. Using the
  semi-empirical Holweger \& M\"uller (1974) model atmosphere, we
  determined the lead non-LTE abundance for the Sun to be
  $\eps{Pb,\odot}$ = 2.09.  We revised the Pb and Eu abundances of the
  two strongly r-process enhanced stars CS~31082-001 and
  HE~1523-0901 and the Roederer et al. (2010) metal-poor stellar
  sample. Our new results provide strong evidence for universal
  Pb-to-Eu relative r-process yields during course of the evolution of
  the Galaxy.
  The stars in the $-2.3 <$ [Fe/H] $< -1.4$ metallicity range have, on
  average, 0.51~dex higher Pb/Eu abundance ratios compared with that
  of the strongly r-process enhanced stars. We conclude that the
  s-process production of lead started as early as the time when
  Galactic metallicity had grown to ${\rm [Fe/H]} = -2.3$.  The
  average Pb/Eu abundance ratio of the mildly metal-poor stars, with
  $-1.4 \le$ [Fe/H] $\le -0.59$, is very close to the corresponding
  Solar System value, in line with the theoretical predictions of
  Travaglio et al. (2001) that AGB stars with [Fe/H] $\simeq -1$
  provided the largest contribution to the solar abundance of s-nuclei
  of lead.  The departures from LTE for \ion{Th}{ii} are caused by the
  pumping transitions from the low-excitation levels, with \Eexc\ $<$
  1~eV. Non-LTE leads to weakened \ion{Th}{ii} lines and positive
  abundance corrections. Overall, the abundance correction does not
  exceed 0.2\,dex when collisions with \ion{H}{i} atoms are taken into
  account in statistical equilibrium calculations. } {}

\keywords{Line: formation -- Nuclear reactions, nucleosynthesis,
  abundances -- Sun: abundances -- Stars: abundances -- Stars:
  atmospheres}

\titlerunning{Non-LTE effects on stellar Pb and Th abundance
  determinations} 

\authorrunning{Mashonkina et al.}

\maketitle

\section{Introduction}

Lead ($Z = 82$), thorium ($Z = 90$), and uranium ($Z = 92$) are the
three heaviest elements observed in metal-poor (MP) stars, with
[Fe/H]\footnote{In the classical notation, where [X/H] = $\log(N_{\rm
    X}/N_{\rm H})_{star} - \log(N_{\rm X}/N_{\rm H})_{Sun}$.} $<
-1$. Only one detection of bismuth ($Z = 83$) has ever been made in an
MP star \citep{2005ApJ...627L.145I}, because the only suitable line,
\ion{Bi}{i} 3067\,\AA, lies in a very crowded spectral region in the
near-UV that is difficult to access. All nuclei between $^{209}$Bi and
$^{244}$Pu and which exist in nature, are radioactive. Among
them, only Th and U have long-lived isotopes with half-lives longer
than 1~Gyr, $t_{1/2}$($^{232}$Th) $\simeq$ 14~Gyr and
$t_{1/2}$($^{238}$U) $\simeq$ 4.5~Gyr, and can be detected in spectra
of MP stars.

The knowledge of stellar Pb abundances provides a better understanding
of the production mechanisms of lead during the evolution of the
Galaxy. Lead is produced by two neutron-capture processes on different
timescales, i.e., in the slow (s) process occuring during thermally
pulsing asymptotic giant branch (AGB) phase of intermediate-mass
(2-8\,M$_{\odot}$) stars (see for example,
\citealt{1998ApJ...497..388G}) and the rapid (r) process. The
astrophysical site at which the r-process operates has not yet been
identified. Given that the heavy elements, such as Ba and Eu, are
detected in very metal-poor (old) stars, supernovae with progenitors
of 8-10\,M$_{\odot}$ are the most promising candidates for r-process
enrichment in the early Galaxy. See, for example, the pioneering
review of \citet{1978SSRv...21..639H} and also
\citet{2004PhT....57j..47C} for further discussion on the
r-process. Theoretical studies of the s-process showed that in
low-metallicity environment, heavy s-nuclei like those of Pb are more
efficiently produced compared to lighter nuclei due to the higher
ratio of free neutrons to Fe-peak seeds, and also due to an extended
period of s-process nucleosynthesis
\citep{1998ApJ...497..388G,2001ApJ...549..346T}. It was predicted that
the AGB stars with [Fe/H] $\simeq -1$ made the greatest contribution
to the solar abundance of s-nuclei of lead. According to
\citet{2001ApJ...549..346T}, the s-process fraction of the solar Pb
amounts to 81\,\%.

The largest data set on lead abundances of MP stars was compiled by
\citet{Roederer2010}.
They found that the stellar Pb/Eu abundance ratios form a plateau in
the metallicity range $-2.3 <$ [Fe/H] $< -1.4$, and Pb/Eu grows at
higher metallicity. The increase of Pb/Eu at [Fe/H] $> -1.4$ suggests
a growing contribution to the stellar lead abundance from the
s-process opening in AGB stars. \citet{Roederer2010} concluded that
s-process material was not widely dispersed until the Galactic
metallicity grew considerably, as high as [Fe/H] $= -1.4$. This also
means that lead in stars with lower metallicity should be of pure
r-process origin.  It is worth noting that despite the large scatter
in the Pb/Eu abundance ratios in the \citet{Roederer2010} stars with
$-2.3 <$ [Fe/H] $< -1.4$, the ratios are, on average, 0.85~dex higher
compared with the corresponding value in the strongly r-process
enhanced star CS~31082-001 \citep{plez_lead}. With [Eu/Fe] = 1.6
\citep{hill2002}, this star experienced enrichment from, probably, a
single r-process source. Following the suggestion of \citet{HERESI},
we hereafter refer to stars having $\mathrm{[Eu/Fe]} > +1$ and
$\mathrm{[Ba/Eu]} < 0$ as r-II stars. The heavy element abundance
ratios of each r-II star are expected to reflect the relative yields
of the r-process. What can be the reason of the large discrepancy in
Pb/Eu between CS~31082-001 and the $-2.3 <$ [Fe/H] $< -1.4$ stars?

In the visual spectra of MP stars, lead is represented by the lines of
\ion{Pb}{i}. Due to a relatively low ionization energy of 7.43~eV,
lead is mostly ionized in the stellar atmospheres with effective
temperature hotter than $\Teff$ = 4000~K, and the number density of
neutral lead can easily deviate from the thermodynamic equilibrium
(TE) population, owing to deviations of the mean intensity of ionizing
radiation from the Planck function.  All previous stellar Pb analyses
have been made under the local thermodynamic equilibrium (LTE)
assumption, and the obtained results should be checked for any
departures from LTE.

The detection of thorium permits, in principle, a nucleo-chronometric
age estimate of the star by comparing the observed Th-to-stable
neutron-capture element-abundance ratios with the corresponding
initial values provided by the r-process event just prior to the time
when the star was born \citep[see, for
  example,][]{sneden1996,cayrel2001,he1523}. Given that the half-life
of $^{232}$Th is long, stellar thorium has to be measured very
precisely, with abundance errors of no more than 0.05~dex, to provide
less than 2.3~Gyr uncertainty in the star's age.

This study aimed to improve the calculation of the \ion{Pb}{i} and
\ion{Th}{ii} lines in stellar atmospheres based on the non-local
thermodynamic equilibrium (non-LTE) line formation and to evaluate
systematic abundance errors caused by the simplified LTE line
formation treatment.  We constructed, for the first time, model atoms
for \ion{Pb}{i} and \ion{Th}{ii}. The non-LTE method and the
departures from LTE for \ion{Pb}{i} are described in
Sect.\,\ref{sec:nlte_pb}. Our theoretical results were used to
determine the solar lead abundance (Sect.\,\ref{sec:sun}) and to
revise the Pb abundances of the \citet{Roederer2010} stellar sample
(Sect.\,\ref{sec:pb_stars}). Having added the r-II star
  HE~1523-0901 \citep{he1523}, we found that the Pb/Eu abundance
ratios of the two r-II stars are well reproduced by the waiting-point
r-process model as presented by \citet{Roederer2009}, while an
additional source of lead, most probably, the s-process in AGB stars,
is needed to explain the Pb abundances of the stars with [Fe/H] $>
-2.3$. The non-LTE calculations for \ion{Th}{ii} are presented in
Sect.\,\ref{sec:nlte_th}.  Our conclusions are given in
Sect.~\ref{Conclusions}.

\section{Non-LTE calculations for \ion{Pb}{i}}\label{sec:nlte_pb}

To solve the coupled radiative transfer and statistical equilibrium
(SE) equations, we used a revised version of the DETAIL program
\citep{detail} based on the accelerated lambda iteration (ALI) method
as described by \citet{rh91,rh92}. The update was presented by
\citet{mash_fe}. The non-LTE calculations were performed with the
MARCS model atmospheres \citep{Gustafssonetal:2008}\footnote{\tt
  http://marcs.astro.uu.se}.

\subsection{\ion{Pb}{i} atomic structure calculations}\label{sec:pb1_calc}

The atomic structure of neutral lead has been well studied with
laboratory measurements.  The ground level of \ion{Pb}{i} is
\eu{6s^26p^2}{3}{P}{}{0}, and the ionization of the 6p electron gives
the ionic core (\ion{Pb}{ii}) 6s$^2$6p, which splits into two levels,
$^2$P$^\circ_{1/2}$ (lower ionization limit 59819.57~cm$^{-1}$) and
$^2$P$^\circ_{3/2}$ (upper ionization limit 73900.64~cm$^{-1}$) in the
LS coupling scheme. The \ion{Pb}{i} energy structure more closely
approaches the $JJ$ than the $LS$ coupling.  Thus, the ground
configuration 6s$^2$6p$^2$ possesses 5 energy levels which are
designated as \jj{1}{1}{0}, \jj{1}{3}{1}, \jj{3}{1}{2}, \jj{3}{3}{2},
and \jj{3}{3}{0}. The energy separation between the 6s$^2$6p$^2$
levels is remarkably large, for example, the second and the fifth
level have an excitation energy of \Eexc\,= 0.97~eV and 3.65~eV,
respectively. Almost all the remaining known energy levels of
\ion{Pb}{i} belong to the 6s$^2$6pnl electronic
configurations. Hereafter, 6s$^2$ is omitted in the configuration
designations.

The early data on the \ion{Pb}{i} energy levels obtained before 1958
were summarized by \citet{Moore}. After that, the analysis was greatly
extended using classical emission \citep{wood_lev} and absorption
\citep{brown_lev} spectroscopy as well as selective laser excitation
(the papers relevant to our study are cited below). The measurements
of long Rydberg series terminating on the first ionization limit
provided many odd parity levels of the 6pns and 6pnd \citep[][up to n
  = 59 and n = 77, respectively]{brown_lev}, and 6png (n = 5-9 in
\citet{wood_lev} and n = 21-47 in \citet{1994PhRvA..49..745D})
configurations. Even parity levels of the 6pnp (n = 7-60) and 6pnf (n
= 5-56) configurations were measured by \citet{hasegawa_lev} a total
of 205 energy levels with a total angular momentum $J$ = 0, 1, and
2. Information on several hundred autoionizing levels above the first
ionization limit is available \citep[see][and references
  therein]{2005EPJD...32..271A}.  

The data on transition probabilities in \ion{Pb}{i} available in the
literature \citep{pb_gf_biemont,FW96} are incomplete, and we rely on
oscillator strengths computed in this study.
The atomic structure of \ion{Pb}{i} was calculated using the Cowan
code \citep{Cowan_code}. The 22 even parity interacting configurations
6p$^2$ + 6pnp (7 $\le$n$\le$ 15) + 6pnf (5 $\le$n$\le$ 12) + 6p6h +
6p7h + 6s6p$^2$7s + 6p$^4$ were included in the energy matrix, while
the odd set was restricted to 24 configurations, i.e., 6pns (7
$\le$n$\le$ 12) + 6pnd (6 $\le$n$\le$ 12) + 6png (5 $\le$n$\le$ 12) +
6sp$^2$7p + 6s6p$^3$ + 5d$^9$6s$^2$6p$^3$. The calculations generally
follow the scheme used by \citet{pb_gf_biemont}. During the fitting of
the calculated energy levels to the experimental ones, the number of
varying energy parameters was restricted as much as possible. All the
parameters describing the configurations in the 6pnl series for
different n were 
varied collectively keeping the ratios of the corresponding {\em ab initio} values.
The exchange Slater integrals G$^k$ within the configurations
were also varied in a similar way. The electrostatic
parameters including the configuration interaction parameters, which
were not optimized in the fitting procedure, were scaled down by a
factor of 0.80, while the spin-orbit integrals were fixed at their
{\em ab initio} values. As a result, the 67 even parity levels were
described by the 25 parameters with a standard deviation of
73~cm$^{-1}$ and the 51 odd parity levels by 21 parameters with a
standard deviation of 74~cm$^{-1}$. Our atomic structure calculations complemented the
system of known energy levels in the region above 6.87~eV with the 20
even parity levels of the 6pnf (n = 9-12, $J$ = 3 and 4) and 6pnh (n =
6, 7) configurations and 29 odd parity levels belonging mostly to the
6png (n = 5-12) configurations.

\begin{figure}
\flushleft 
\hspace{-5mm}
  \resizebox{88mm}{!}{\includegraphics{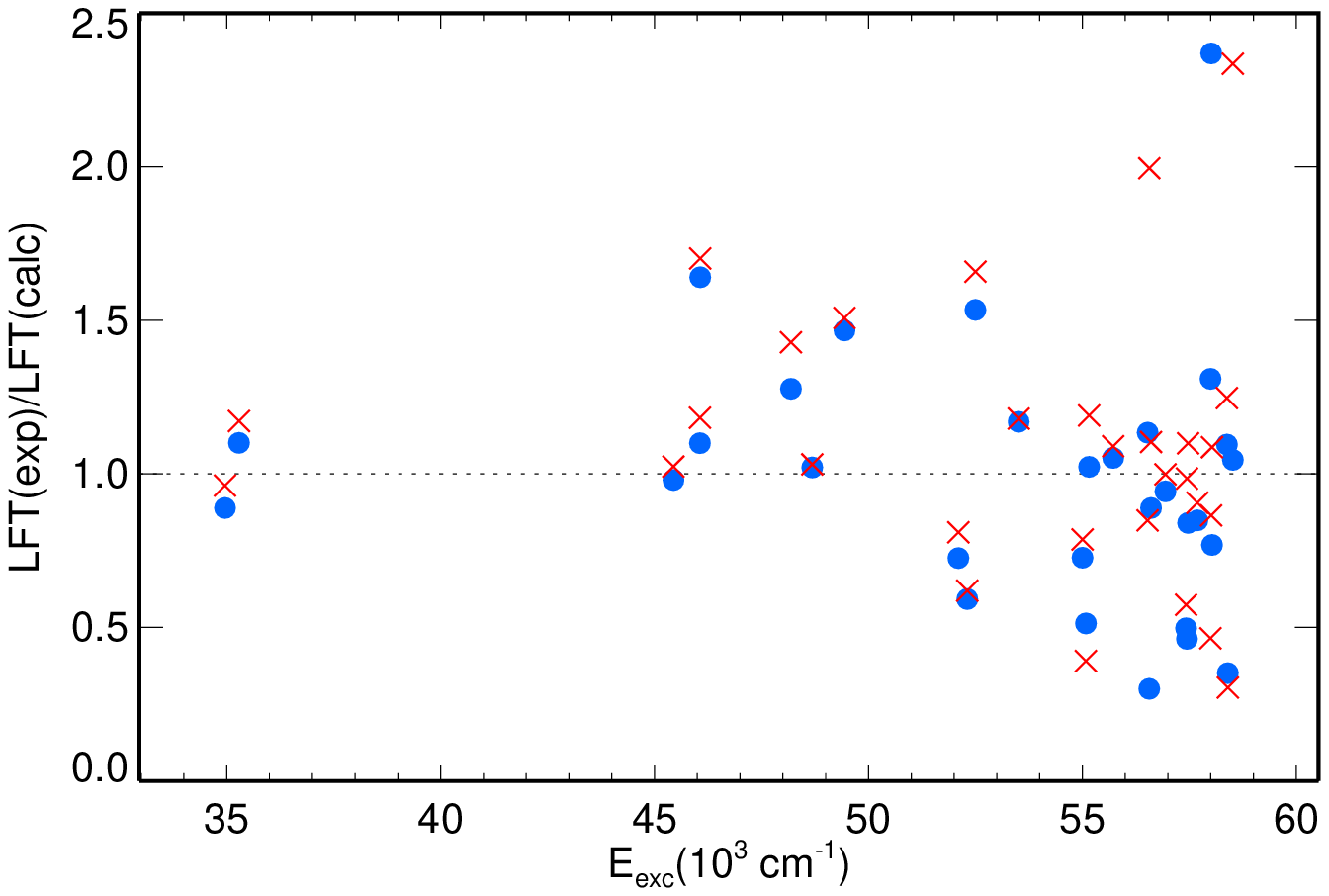}}
\caption{\label{lifetime_comp} Measured \citep{pb_gf_biemont} to
  calculated \ion{Pb}{i} lifetime ratios for the 6pns (7 $\le$n$\le$
  12) and 6pnd (6 $\le$n$\le$ 13) levels, mostly with $J$ = 1 and
  2. Open circles and crosses correspond to our and the
  \citet{pb_gf_biemont} calculations, respectively. }
\end{figure}

The wavefunctions obtained after the fitting of energy levels were
used for the calculations of transition probabilities. The dipole
transition integrals for all transitions were kept at {\em ab initio}
Hartree-Fock values. The accuracy of our calculations was estimated by
the comparison with the experimental data for common levels.
Figure\,\ref{lifetime_comp} shows that the difference in lifetime
between our calculations and the measurements of \citet{pb_gf_biemont}
is mostly within 75\,\%, with the only outlier being the measured to
calculated lifetime ratio of about 2.3. On average, the predicted
lifetimes approach to the measured values.  It should be noted that
the lifetimes of the highest 6pnd levels with $J$ = 2 are critically
dependent on a position of strongly interacting
\eu{6s6p^3}{5}{S}{\circ}{2} level. It seems that the only
$^3$P$^\circ_1$ level of the 6s6p$^3$ configuration is firmly located
at \Eexc\ = 85870~cm$^{-1}$
\citep{1986PhRvA..34.4511K,1990PhyS...41...38M}. \citet{1994PhRvA..49..745D}
estimated the position of the \eu{6s6p^3}{3}{D}{\circ}{1} level at
\Eexc\ = 68943~cm$^{-1}$ from analysis of the interactions in the 6pns
and 6pnd series. They also established that ``the state
\eu{6s6p^3}{5}{S}{\circ}{2} does not occur as a dominant component in
any of the electronic levels but it mixed into many odd levels''. Its
unperturbed value should be around 59300~cm$^{-1}$. To predict the
levels of the 6s6p$^3$ configuration, the same scaling of \textit{ab  initio} values 
was applied to the parameters within the 6p$^3$
subshell as that for the 6s$^2$6p$^2$ configuration. 
The average energy and the G$^1$(6s,6p) parameter were adjusted to fit the
energies of the $^3$P$^\circ_1$ and $^3$D$^\circ_1$ levels. The energy
of the \eu{6s6p^3}{5}{S}{\circ}{2} level appeared to be
58700~cm$^{-1}$, in quite good agreement with the estimates of
\citet{1994PhRvA..49..745D}.  Similar consistency between  theory
and measurements was found for the even parity 6pnp (7 $\le$n$\le$ 14)
and 6pnf (6 $\le$n$\le$ 8) levels when using the measured lifetimes of
\citet{2001JPhB...34.3501L}.

Figure\,\ref{lifetime_comp} shows also the measured to calculated
lifetime ratios from the calculations of \citet{pb_gf_biemont}, who
took into account the core polarization effect that is expected to
give rise to the precision of the predicted lifetimes. As can be seen,
our results obtained without taking into account the core
polarization agree very well with the calculations of
\citet{pb_gf_biemont}. Therefore, $gf-$values calculated in this study
are expected to have the same accuracy as the values in
\citet{pb_gf_biemont} for all the transitions in \ion{Pb}{i}.

\begin{figure}
\flushleft 
\hspace{-15mm}
  \resizebox{88mm}{!}{\includegraphics{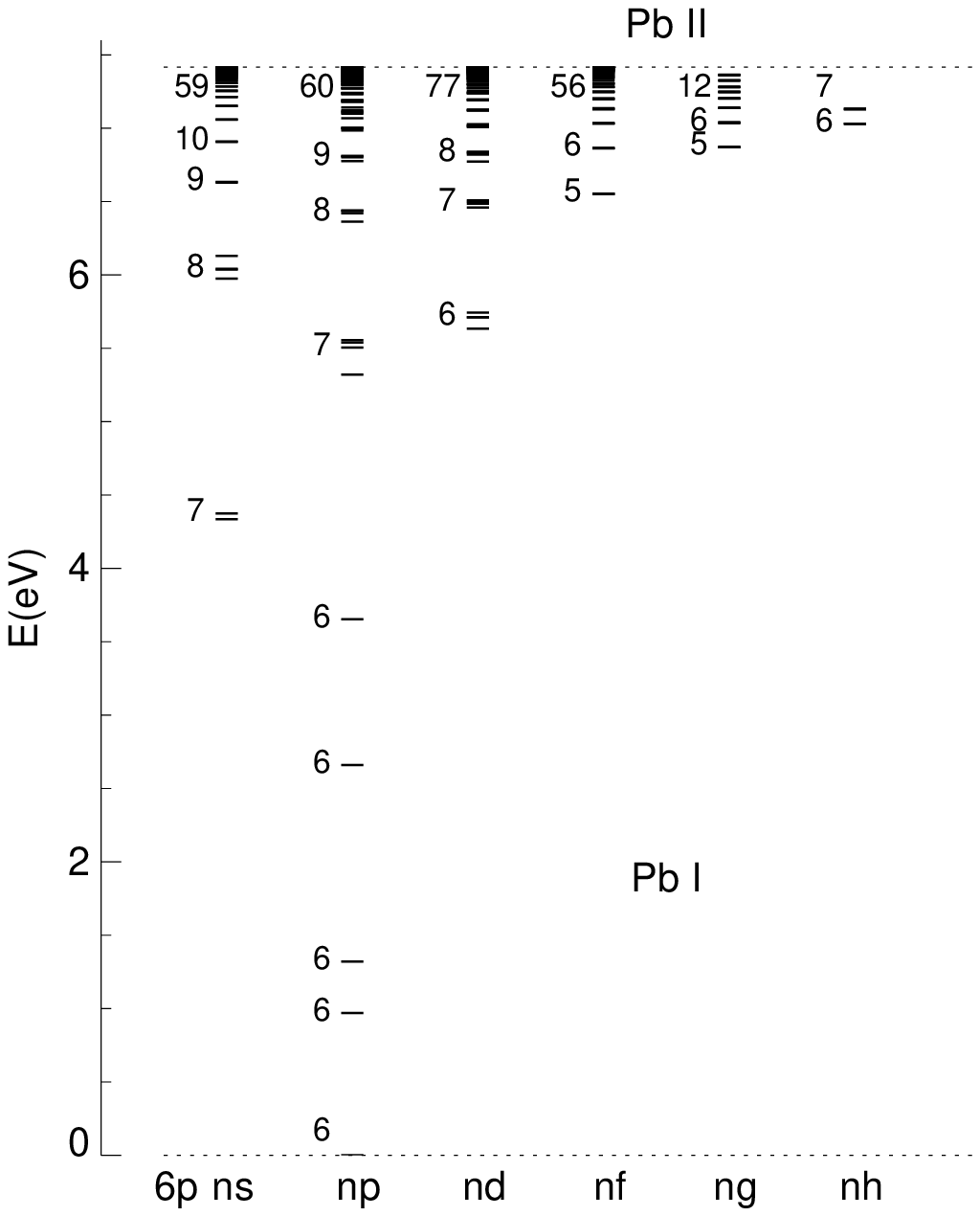}}
\caption{\label{Pb_atom} Measured and predicted energy levels of
  \ion{Pb}{i} that form our final Pb model atom.}
\end{figure}

\subsection{Lead model atom}

{\it Energy levels.} To construct the model atom, we used all the
measured and predicted levels in \ion{Pb}{i}. They are shown in
Fig.~\ref{Pb_atom}.  The high-excitation (\Eexc\, $\ge$ 6.04~eV)
levels with common parity and energy separation of smaller than
0.01~eV were combined into a single level. The final model atom is
fairly complete. It includes 97 levels of \ion{Pb}{i} and the ground
state \eu{6s^26p}{2}{P}{\circ}{1/2} of \ion{Pb}{ii}. The second
sublevel of the ground term, $^2$P$^\circ_{3/2}$ (\Eexc\,= 1.75~eV),
was taken into account only in the number conservation equation. The
excited electronic configurations of \ion{Pb}{ii} produce levels with
an excitation energy of more than 7.2~eV, and were thus ignored.

{\it Radiative rates.} For 141 transitions in \ion{Pb}{i} that connect
the 6p$^2$ and 6p7p levels to the 6pns and 6pnd (n$\le 13$) ones, we
employed accurate $gf-$values based on measured natural radiative
lifetimes and theoretical branching ratios from \citet{pb_gf_biemont}
and \citet{FW96}. For the remaining majority of transitions, we used
oscillator strengths computed in this study. In total, 1\,954
radiative bound-bound ($b-b$) transitions in \ion{Pb}{i} were included
in our SE calculations.

For the \ion{Pb}{i} ground state, we applied the photoionization
cross-sections $\sigma_{\rm ph}$ measured by \citet{ph_ion1} from the
ionization threshold at 1671 to 1470\,\AA. For the remaining levels,
$\sigma_{\rm ph}$ were computed using the hydrogen approximation
formula with n = n$_{\rm eff}$, where n$_{\rm eff} = Z \sqrt{\chi_{\rm
    H}/\chi_i}$ is the effective principal quantum number for the
level with the ionization energy $\chi_i$. Here, $Z$ = 1 and
$\chi_{\rm H}$ is the hydrogen ionization energy.  Such a choice was
justified by the comparison of the experimental threshold
cross-section $\sigma_{{\rm thr},exp}$ = 10\,Mb for the \ion{Pb}{i}
ground state with the cross-sections calculated with various n. The
use of n = 6 results in the too low hydrogen approximation
cross-section $\sigma_{{\rm thr},hyd}$ = 0.0007\,Mb. With n$_{\rm
  eff}$ = 1.35, we computed $\sigma_{{\rm thr},hyd}$ = 1.3\,Mb, which
is much closer to the experimental data, although it is about a factor
of 10 lower.

{\it Collisional rates.}  All levels in our model atom are coupled via
collisional excitation and ionization by electrons and by neutral
hydrogen atoms. The calculations of collisional rates rely on
theoretical approximations. For electron-impact excitation, we used
the formula of \citet{Reg1962} for the allowed transitions and we
assumed that the effective collision strength $\Upsilon$ = 1 for the
forbidden transitions. Electron-impact ionization cross-sections were
computed according to \citet{Drawin61}.

For collisions with \ion{H}{i} atoms, we employed the classical Drawin
formula as described by \citet{Steenbock1984}. Over the past few
decades, this formula has been criticized for significantly
overestimating the collision rates \citep[see][and references
  therein]{barklem2011}. At the same time, the need for a thermalizing
process not involving electrons in the atmospheres of, in particular,
very metal-poor stars, was indicated by many non-LTE spectral line
formation studies \citep[see][and references
  therein]{mash_fe}. Because no accurate calculations of either
inelastic collisions of lead with neutral hydrogen atoms or other
types of processes are available, we simulate an additional source of
thermalization in the atmospheres of cool stars by using parametrized
\ion{H}{i} collisions. The Drawin formula can only be applied to
allowed $b-b$ and $b-f$ transitions. For the forbidden transitions, a
simple relation between hydrogen and electron collision rates, $C_H =
C_e \sqrt{(m_e/m_H)} N_H/N_e$, was used following
\citet{Takeda95}. The SE calculations were performed with various
efficiencies of collisions with \ion{H}{i} atoms by applying a scaling
factor of \kH\ = 0 (no hydrogen collisions), 0.1, and 1.

We discuss below the effects of the uncertainty in the adopted
photoionization cross-sections and collisional rates on our final
results.

\subsection{Departures from LTE for \ion{Pb}{i}}

\begin{figure}
\flushleft 
\hspace{-5mm}
\resizebox{88mm}{!}{\includegraphics{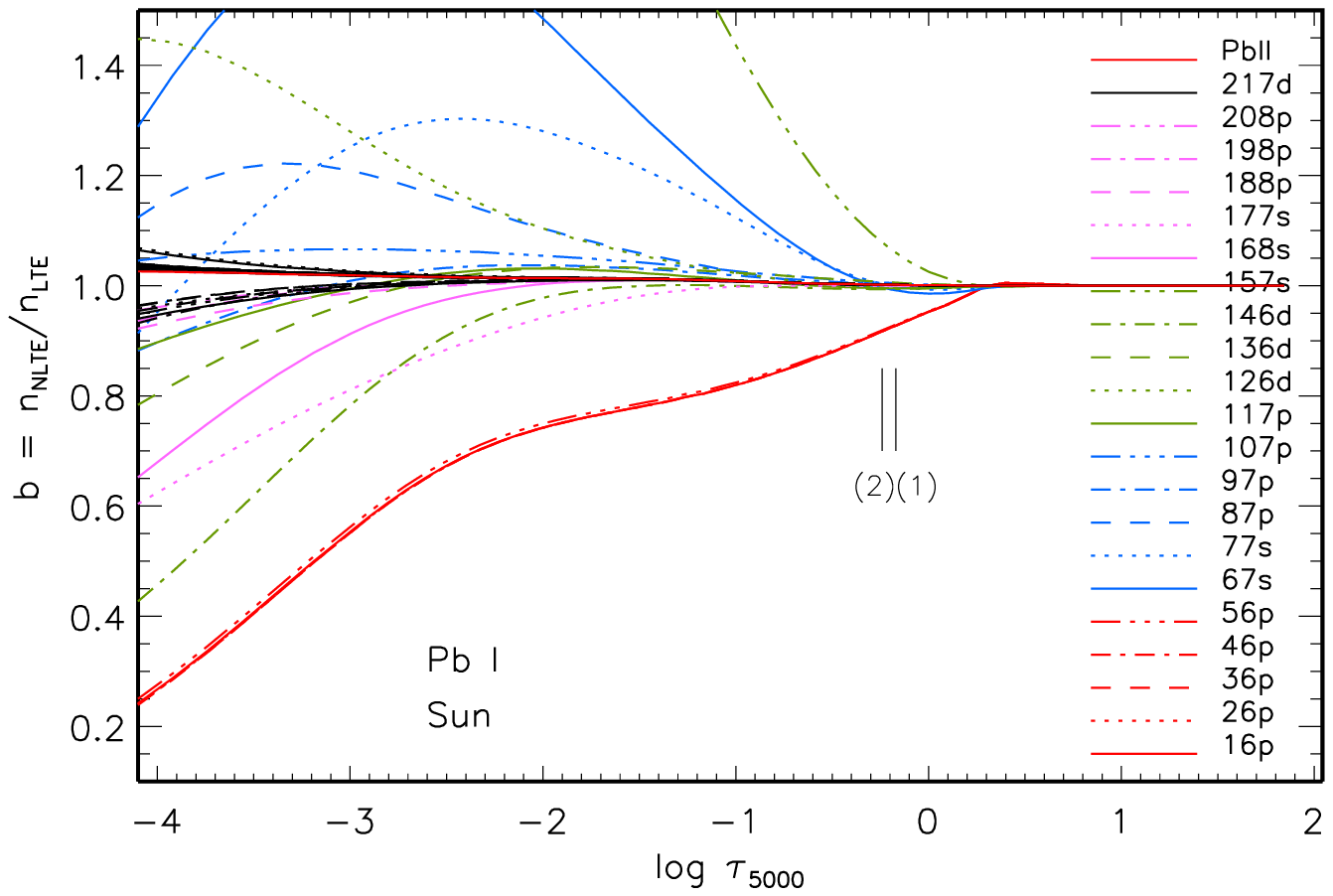}}

\hspace{-5mm}
\resizebox{88mm}{!}{\includegraphics{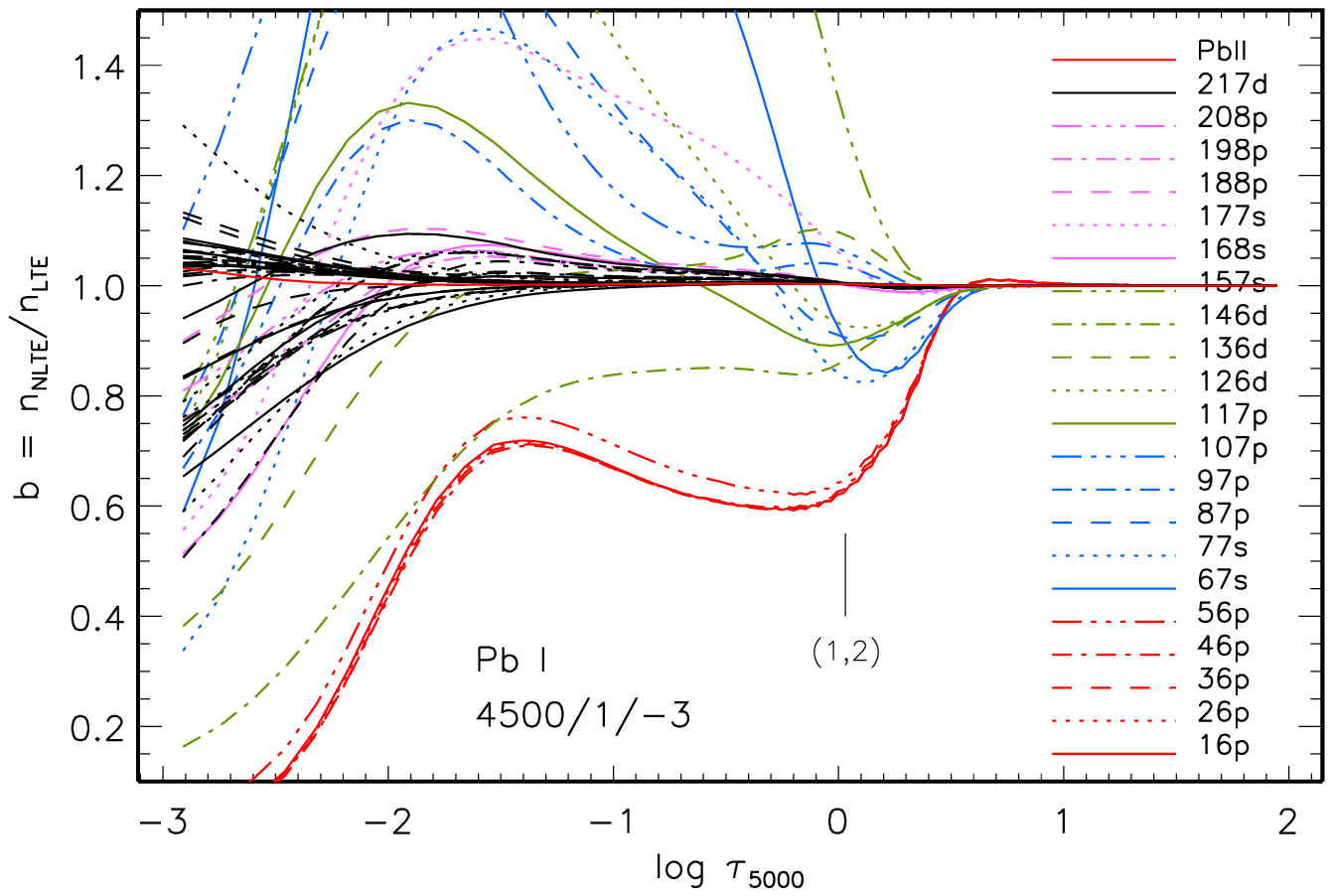}}
\caption[]{Departure coefficients, $b$, for the lowest 60 levels of
  \ion{Pb}{i} and the ground state of \ion{Pb}{ii} as a function of
  $\log \tau_{5000}$ in the model atmospheres 5780/4.44/0 (top panel)
  and 4500/1.0/$-3$ (bottom panel). The first 21 levels are quoted in
  the right part of each panel. For all the higher excitation levels,
  their behaviour is very similar to that of the 188p to 217d levels,
  with $b \simeq 1$ in the solar and VMP models inside $\log
  \tau_{5000}$ = $-3$ and $-1$, respectively. \kH\ = 0.1 was used
  throughout. The vertical lines indicate the locations of line core
  formation depths for \ion{Pb}{i} 4057 (1) and 3683\,\AA (2). See
  text for more details.} \label{Fig:bf_pb}
\end{figure}

We first inspect the statistical equilibrium of lead in the two model
atmospheres representing the solar atmosphere, with $\Teff$/$\log
g$/[Fe/H] = 5780/4.44/0, and the atmosphere of a typical very
metal-poor (VMP) cool giant (4500/1.0/$-3$) from the calculations with
\kH\ = 0.1. The departure coefficients, $b_i = n_i^{\rm NLTE}/n_i^{\rm
  LTE}$, for the \ion{Pb}{i} and \ion{Pb}{ii} levels are shown in
Fig.\,\ref{Fig:bf_pb}. Here, $n_i^{\rm NLTE}$ and $n_i^{\rm LTE}$ are
the statistical equilibrium and TE number densities, respectively. In
these computations, the Pb abundance was adopted at $\eps{Pb} = 1.92$
in the solar model and at $\eps{Pb} = -0.08$ ([Pb/Fe] = +1) in the VMP
model.  In each model, lead is almost completely ionized, such that
the fraction of \ion{Pb}{i} nowhere exceeds 4\%\ in the solar
atmosphere and 0.6\%\ in the cool VMP atmosphere. This explains why
the non-LTE mechanisms for \ion{Pb}{i} are similar in the two
models. We summarize our findings as follows.

1. In the atmosphere outside $\log \tau_{5000}$ = $+0.2$ and $+0.5$
for the solar and VMP models, respectively, the most populated 6p$^2$
levels of \ion{Pb}{i} are underpopulated relative to their TE number
densities and they are close together. The main non-LTE mechanism is
the ultraviolet (UV) overionization caused by superthermal radiation
of a non-local origin below the thresholds at $\lambda_{thr}$ = 1923,
2033, and 2606\,\AA\ for the three excited 6p$^2$ levels.

2. The populations of the 6p7s levels labeled in the model atom as 67s
and 77s are controlled by the two mechanisms. The UV overionization
($\lambda_{thr}$ = 4022 and 4076\,\AA) tends to drain the populations
far inside the atmosphere. The competing process is radiative pumping
of the strong 6p$^2$ -- 6p7s transitions \jj{1}{1}{0} --
\jj{1}{1}{1}$^\circ$ (16p - 77s in our denotations, which produces the
line at 2833\,\AA), \jj{1}{3}{1} -- {\jj{1}{1}{0}}$^{\circ}$ (26p -
67s, 3683\,\AA), and \jj{1}{3}{1} -- \jj{1}{1}{1}$^\circ$ (26p - 77s,
3639\,\AA) by the ultraviolet $J_\nu - B_\nu(T_e)$ excess radiation
which tends to produce enhanced excitation of the upper levels in the
atmospheric layers, where the line wing optical depth drops below
1. The net effect is that the 67s and 77s levels are underpopulated
far inside the atmosphere, at $\log \tau_{5000}$ = $-0.7$ to $+0.2$
and $-0.3$ to $+0.2$ in the models 4500/1.0/$-3$ and 5780/4.44/0,
respectively, and they are overpopulated in the higher layers.  No
process seems to compete with radiative pumping of the 26p - 157s
(2476\,\AA) and 36p - 157s (2663\,\AA) transitions resulting in strong
overpopulation of the upper level in the atmosphere outside $\log
\tau_{5000}$ = $+0.2$.

3. The overpopulation of the 6p7s levels is redistributed to the 6p7p
levels through inelastic collisions.

4. All the high-excitation levels with \Eexc\, $\ge$ 6.1~eV (starting
from the 18th level in the model atom) couple thermally to the
\ion{Pb}{ii} ground state.

5. \ion{Pb}{ii} represents the state in which the majority of the
element exists and it preserves, therefore, the TE population.

The formation depth has been specified as the depth where $\tau_\nu
\lambda_\nu^{1/2}$ = 1, following \citet{mihalas78}. Here,
$\lambda_\nu$ is the ratio of the thermal to total absorption
coefficient. Everywhere in the solar atmosphere, $\lambda_\nu \simeq
1$, irrespective of the frequency point. In the cool VMP model,
$\lambda_\nu < 1$ due to significant contribution of Rayleigh
scattering to the total opacity. For example, at the depth point
$\log\tau_{5000} = 0$, $\lambda_\nu \simeq 0.7$ at 4057\,\AA\ and it
decreases towards shorter wavelengths. Even lower $\lambda_\nu$ values
are obtained in the higher atmospheric layers.

The two lines, \ion{Pb}{i} 4057\,\AA\ (6p$^2$ \jj{1}{3}{2} -- 6p7s
{\jj{1}{1}{1}}$^{\circ}$ denoted as 36p -- 77s in our model atom) and
3683\,\AA\ (6p$^2$ \jj{1}{3}{1} -- 6p7s {\jj{1}{1}{0}}$^{\circ}$
denoted as 26p - 67s), and, in many cases, only one of them are used
in stellar Pb abundance analyses. In each model, non-LTE leads to a
weakening of the \ion{Pb}{i} lines compared to their LTE strengths and
thus positive non-LTE abundance corrections. This is due to
overionization ($b_{l} < 1$) and also owing to the line source
function $S_{lu} \simeq b_u/b_l\,B_\nu$ rises ($b_u > b_l$) being
above the Planck function $B_\nu$ in the line-formation layers
(Fig.\,\ref{Fig:bf_pb}). Here, $b_u$ and $b_l$ are the departure
coefficients of the upper and lower levels, respectively.  We found
that the departures from LTE are significantly larger in the VMP model
compared to the solar one. The non-LTE abundance correction,
$\Delta_{\rm NLTE} = \eps{NLTE}-\eps{LTE}$, decreases from $+0.21$
down to $+0.09$\,dex for \ion{Pb}{i} 3683\,\AA\ in the solar model
when \kH\ moves from 0 to 1. Similar values, $+0.19$ and $+0.11$\,dex,
were obtained for \ion{Pb}{i} 4057\,\AA. In the 4500/1/$-3$ model,
$\Delta_{\rm NLTE}$(\ion{Pb}{i} 4057\,\AA) $= +0.49$ and $+0.38$~dex
when \kH\ = 0 and 1, respectively.

\begin{table}
 \centering
 \caption{\label{pb_eu_corr} Non-LTE abundance corrections (dex) for the \ion{Pb}{i} 4057\,{\AA} and \ion{Eu}{ii} 4129\,{\AA} lines from the calculations with \kH\,= 0.1.}
  \begin{tabular}{ccccc}
   \hline\noalign{\smallskip}
$\Teff$ & $\log g$ & [Fe/H]  & \multicolumn{2}{c}{$\Delta_{\rm NLTE}$} \\
\noalign{\smallskip} \cline{4-5} \noalign{\smallskip}
 & &    &      \ion{Pb}{i} & \ion{Eu}{ii} \\
   \hline\noalign{\smallskip}
5780 & 4.44 & 0  & 0.16 & 0.03 \\
\multicolumn{5}{c}{[Pb/Fe] = 0.5$^1$, [Eu/Fe] = 0.7$^1$} \\
4750 & 1.5 & --1 & 0.26 & 0.06 \\
5000 & 3.0 & --1 & 0.32 & 0.06 \\
5000 & 4.0 & --1 & 0.28 & 0.05 \\
5000 & 4.5 & --1 & 0.22 & 0.04 \\
5500 & 4.0 & --1 & 0.31 & 0.05 \\
5500 & 4.5 & --1 & 0.27 & 0.05 \\
4000 & 0.0 & --2 & 0.38 & 0.12 \\
4000 & 0.5 & --2 & 0.32 & 0.08 \\
4250 & 0.5 & --2 & 0.33 & 0.07 \\
4250 & 1.0 & --2 & 0.30 & 0.06 \\
4500 & 1.0 & --2 & 0.30 & 0.06 \\
4500 & 1.5 & --2 & 0.32 & 0.06 \\
4750 & 1.5 & --2 & 0.37 & 0.07 \\
4750 & 2.0 & --2 & 0.39 & 0.07 \\
5250 & 2.5 & --2 & 0.52 & 0.10 \\
\multicolumn{5}{c}{[Pb/Fe] = 1.0$^1$, [Eu/Fe] = 1.8$^1$} \\
4500 & 1.0 & --3 & 0.41 & 0.04 \\ 
4500 & 1.5 & --3 & 0.42 & 0.04 \\ 
4750 & 1.0 & --3 & 0.49 & 0.06 \\ 
4750 & 1.5 & --3 & 0.53 & 0.06 \\ 
5000 & 1.5 & --3 & 0.62 & 0.07 \\
\noalign{\smallskip} \hline \noalign{\smallskip}
\multicolumn{5}{l}{ \ $^1$ Abundances used in non-LTE calculations.} \\
\end{tabular}
\end{table}

Using \kH\ = 0.1, we computed $\Delta_{\rm NLTE}$ for the grid of model atmospheres
 characteristic of the Galactic halo stars. 
Table~\ref{pb_eu_corr} presents the resulting values for
\ion{Pb}{i} 4057\,\AA. It can be seen that the non-LTE correction
grows toward lower metallicity from $\Delta_{\rm NLTE}$ = 0.16~dex in
the solar model up to $\Delta_{\rm NLTE}$ = 0.62~dex in the
5000/1.50/$-3$ one. For a given metallicity, $\Delta_{\rm NLTE}$
increases with increasing $\Teff$ and decreasing $\log g$.

Besides different prescriptions of the collisions with hydrogen atoms,
two different test calculations were performed for the 4500/1/$-3$
model to assess the influence of crucial atomic data on final
results. Here, we used \kH\,= 0.1. First, we varied photoionization
cross-sections by employing the principal quantum number n instead of
n$_{\rm eff}$ in the hydrogen approximation formula. For the 6p$^2$ and
6p7s levels, this resulted in the reduction of $\sigma_{ph}$ by a
factor of 150 to 70. However, the effect on $\Delta_{\rm NLTE}$ for
\ion{Pb}{i} 4057\,{\AA} was found to be minor, of less than 0.01~dex,
because the departures from LTE for this line are controlled by the
$b-b$ radiative transitions between the 6p$^2$ and 6p7s levels.

Second, we checked the atomic model, where \ion{H}{i} collisions were
taken into account for the allowed transitions, but were neglected for
the forbidden ones. As expected, this resulted in strengthening the
departures from LTE. For \ion{Pb}{i} 4057\,{\AA}, $\Delta_{\rm NLTE}$
grew from 0.41~dex up to 0.53~dex. Thus, poor knowledge of collisions
with \ion{H}{i} atoms is the main source of the uncertainty in the
calculated non-LTE abundance corrections for the \ion{Pb}{i} lines.

\section{Solar lead abundance}\label{sec:sun}

As a test and first application of the \ion{Pb}{i} model atom, we
determined the lead abundance for the Sun. We used solar central
intensity observations taken from the Kitt Peak Solar Atlas of
\citet{brault1972kitt}. The only useful line of \ion{Pb}{i} at
3683.46\,\AA\ lies in a rather crowded spectral region, and the
element abundance was derived applying the line-profile fitting method
with the {\sc SIU} program \citep{Reetz}.  We employed the
semi-empirical solar model atmosphere of \citet[][hereafter,
  HM74]{HM74} and the theoretical MARCS model
\citep{Gustafssonetal:2008} with $\Teff$ = 5780~K and $\log g$ =
4.44. A depth-independent microturbulence of 0.9\,\kms\ was adopted.

\begin{figure}
\flushleft 
\hspace{-5mm}
  \resizebox{88mm}{!}{\includegraphics{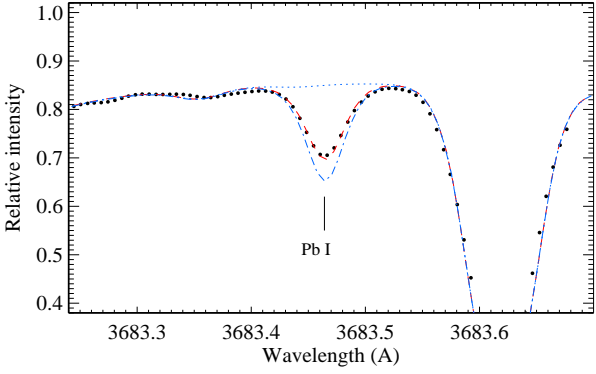}}
\caption{\label{solar3683} Synthetic non-LTE (\kH\ = 0.1, dashed
curve) and LTE (dash-dotted curve) disk-center intensity profiles of \ion{Pb}{i} 3683.46\,\AA\ computed with the HM74 model atmosphere and $\eps{Pb}=2.09$ compared with the observed spectrum of the \citet{brault1972kitt} solar atlas (bold dots). Dotted curve corresponds to the atmosphere without lead. 
}
\end{figure}

The continuum level in the spectral range 3678-3688\,\AA\ was fixed
using an atomic line list from the {\sc VALD} database \citep{vald}
and a molecular line list from \citet{Kur94a}.  Our synthetic
intensity profiles were convolved with a radial-tangential
macroturbulence profile of $\Vmac$ = 2.5\,\kms.
Figure\,\ref{solar3683} illustrates the fit of the 3683\,\AA\ blend,
in particular, \ion{Pb}{i} 3683.46\,\AA\ and the strong \ion{Fe}{i}
3683.612 and 3683.629\,\AA\ lines.  For \ion{Pb}{i} 3683.46\,\AA,
$\log gf = -0.52$ \citep{pb_gf_biemont} was employed throughout.

Under the LTE assumption, we obtained $\eps{Pb,LTE}$ = 1.96 with the
HM74 model and $\eps{Pb,LTE}$ = 1.85 with the solar MARCS model. It
should be stressed that the solar photosphere lead LTE abundance is
lower than the meteoritic one $\eps{Pb,met}$ = 2.04$\pm$0.03
\citep{aspl09} and 2.06$\pm$0.03 \citep{Lodders2009}, independent of
the applied model atmosphere.  The two recent LTE determinations
yielded $\eps{Pb,LTE}$ = 1.75$\pm$0.10 based on the theoretical 3D
model \citep{aspl09} and 2.00$\pm$0.06 based on the HM74 model
\citep{Lodders2009}.

We found that non-LTE decreases the discrepancy between the solar
photosphere and meteoritic Pb abundance. When using the HM74 model,
the non-LTE abundances amount to $\eps{Pb}$ = 2.12, 2.09, and 2.03
from the calculations with \kH\ = 0, 0.1, and 1,
respectively. Figure\,\ref{solar3683} shows the best non-LTE fit for
\kH\ = 0.1.  For the solar MARCS model, the non-LTE abundances are
2.00, 1.97, and 1.92 depending on the used \kH\ value, i.e., for 0,
0.1, and 1, respectively.  We note that, in all cases, the non-LTE
correction calculated for the emergent intensity spectrum is smaller
than that for the flux spectrum. This is because the disk-center
radiation emerges from deeper layers, where the departures from LTE
are weaker.  From the analysis of the \ion{Pb}{i} 3683\,\AA\ line
alone we can not decide about the efficiency of poorly known inelastic
collisions with \ion{H}{i} atoms in SE calculations, and we choose
\kH\,= 0.1 based on our empirical estimates for
\ion{Ca}{i}-\ion{Ca}{ii} and \ion{Fe}{i}-\ion{Fe}{ii}
\citep{mash_ca,mash_fe}. Thus, our final solar lead abundance is
$\eps{Pb,NLTE}$ = 2.09 with the HM74 model atmosphere and
$\eps{Pb,NLTE}$ = 1.97 with the solar MARCS model.

\begin{table*} 
 \centering
 \caption{\label{pb_corr} Non-LTE and LTE abundances of Pb and Eu and
   non-LTE abundance corrections (dex) for the \ion{Pb}{i} 4057\,{\AA}
   and \ion{Eu}{ii} 4129\,{\AA} lines in the \citet{Roederer2010}
   stellar sample.}
  \begin{tabular}{lcccrrccrrc}
   \hline\noalign{\smallskip}
  &    &        &      & \multicolumn{3}{c}{Pb} & \ & \multicolumn{3}{c}{Eu} \\ 
\noalign{\smallskip} \cline{5-7} 
 \cline{9-11} \noalign{\smallskip}
\multicolumn{1}{c}{Star} & $\Teff$ & $\log g$ & [Fe/H] & $\eps{NLTE}^1$ & $\eps{LTE}$ & $\Delta_{\rm NLTE}^1$ & & $\eps{NLTE}^1$ & $\eps{LTE}$ & $\Delta_{\rm NLTE}^1$ \\
\noalign{\smallskip} \hline\noalign{\smallskip}
 BD+01 2916   & 4200 & 0.40 & $-$1.92 & 0.12 & $-$0.20 & 0.32 & & $-$1.14 & $-$1.22 & 0.08 \\
 BD+19 1185   & 5500 & 4.19 & $-$1.09 & 1.06 & 0.77 & 0.29 & & $-$0.18 & $-$0.23 & 0.05 \\
 BD+29 2356   & 4760 & 1.60 & $-$1.59 & 0.65 & 0.35 & 0.30 & & $-$0.62 & $-$0.69 & 0.07 \\
 BD+30 2611   & 4330 & 0.60 & $-$1.50 & 0.92 & 0.56 & 0.36 & & $-$0.29 & $-$0.36 & 0.07 \\
 CS31082-001  & 4825 & 1.50 & $-$2.90 & 0.01 & $-$0.55 & 0.56 & & $-$0.66 & $-$0.72 & 0.06 \\
 G009-036     & 5625 & 4.57 & $-$1.17 & 1.53 & 1.25 & 0.28 & & $-$0.11 & $-$0.16 & 0.05 \\
 G028-043     & 5060 & 4.50 & $-$1.64 & 0.97 & 0.69 & 0.28 & & $-$0.48 & $-$0.53 & 0.05 \\
 G029-025     & 5225 & 4.28 & $-$1.09 & 1.06 & 0.80 & 0.26 & & $-$0.18 & $-$0.23 & 0.05 \\
 G058-025     & 6000 & 4.21 & $-$1.40 & 1.67 & 1.29 & 0.38 & & $-$0.61 & $-$0.66 & 0.05 \\
 G059-001     & 5920 & 3.98 & $-$0.95 & 1.98 & 1.64 & 0.34 & & $-$0.24 & $-$0.29 & 0.05 \\
 G068-003     & 4975 & 3.50 & $-$0.76 & 1.49 & 1.19 & 0.30 & &  0.22 &  0.16 & 0.06 \\
 G095-057A    & 4955 & 4.40 & $-$1.22 & 1.35 & 1.14 & 0.21 & & $-$0.19 & $-$0.23 & 0.04 \\
 G102-020     & 5250 & 4.44 & $-$1.25 & 1.04 & 0.79 & 0.25 & & $-$0.27 & $-$0.32 & 0.05 \\
 G102-027     & 5600 & 3.75 & $-$0.59 & 1.87 & 1.64 & 0.23 & & 0.43 & 0.40 & 0.03 \\
 G113-022     & 5525 & 4.25 & $-$1.18 & 1.49 & 1.19 & 0.30 & & $-$0.08 & $-$0.13 & 0.05 \\
 G122-051     & 4864 & 4.51 & $-$1.43 & 0.54 & 0.34 & 0.20 & & $-$0.23 & $-$0.27 & 0.04 \\
 G123-009     & 5490 & 4.75 & $-$1.25 & 1.40 & 1.13 & 0.27 & & $-$0.16 & $-$0.21 & 0.05 \\
 G126-036     & 5490 & 4.50 & $-$1.06 & 1.94 & 1.67 & 0.27 & & 0.09 & 0.04 & 0.05 \\
 G140-046     & 4980 & 4.42 & $-$1.30 & 1.41 & 1.19 & 0.22 & & $-$0.37 & $-$0.41 & 0.04 \\
 G179-022     & 5080 & 3.20 & $-$1.35 & 0.72 & 0.39 & 0.33 & & $-$0.17 & $-$0.22 & 0.05 \\
 G221-007     & 5020 & 3.37 & $-$0.98 & 1.99 & 1.69 & 0.30 & & $-$0.06 & $-$0.11 & 0.05 \\
 HD 3008      & 4250 & 0.25 & $-$2.08 & $-$0.21 & $-$0.56 & 0.35 & & $-$1.02 & $-$1.09 & 0.07 \\
 HD 6755      & 5100 & 2.93 & $-$1.68 & 0.73 & 0.32 & 0.41 & & $-$0.44 & $-$0.50 & 0.06 \\
 HD 6833      & 4400 & 1.50 & $-$0.85 & 1.93 & 1.69 & 0.24 & & 0.15 & 0.10 & 0.05 \\
 HD 8724      & 4535 & 1.40 & $-$1.91 & 0.32 & 0.00 & 0.32 & & $-$0.80 & $-$0.86 & 0.06 \\
 HD 23798     & 4450 & 1.06 & $-$2.26 & 0.09 & $-$0.21 & 0.30 & & $-$1.30 & $-$1.36 & 0.06 \\
 HD 26297     & 4320 & 1.11 & $-$1.98 & 0.22 & $-$0.11 & 0.33 & & $-$1.16 & $-$1.22 & 0.06 \\
 HD 29574     & 4250 & 0.80 & $-$2.00 & 0.25 & $-$0.06 & 0.31 & & $-$0.57 & $-$0.63 & 0.06 \\
 HD 29907$^2$ & 5500 & 4.64 & $-$1.55 & 1.03 & 0.70 & 0.33 & & $-$0.45 & $-$0.50 & 0.05 \\
 HD 37828     & 4350 & 1.50 & $-$1.62 & 1.14 & 0.77 & 0.37 & & $-$0.48 & $-$0.53 & 0.05 \\
 HD 44007     & 4850 & 2.00 & $-$1.72 & 0.67 & 0.31 & 0.36 & & $-$0.87 & $-$0.94 & 0.07 \\
 HD 63791     & 4675 & 2.00 & $-$1.90 & 0.58 & 0.22 & 0.36 & & $-$0.85 & $-$0.92 & 0.07 \\
 HD 74462     & 4700 & 2.00 & $-$1.52 & 0.79 & 0.49 & 0.30 & & $-$0.32 & $-$0.39 & 0.07 \\
 HD 105755    & 5700 & 3.82 & $-$0.83 & 1.71 & 1.43 & 0.28 & & 0.07 & 0.02 & 0.05 \\
 HD 106516    & 6170 & 4.21 & $-$0.81 & 1.85 & 1.56 & 0.29 & & 0.05 & $-$0.04 & 0.09 \\
 HD 108317    & 5234 & 2.68 & $-$2.18 & 0.69 & 0.17 & 0.52 & & $-$1.22 & $-$1.32 & 0.10 \\
 HD 121135    & 4934 & 1.91 & $-$1.54 & 0.72 & 0.38 & 0.34 & & $-$0.63 & $-$0.70 & 0.07 \\
 HD 122956    & 4508 & 1.55 & $-$1.95 & 0.19 & $-$0.13 & 0.32 & & $-$0.73 & $-$0.79 & 0.06 \\
 HD 141531    & 4360 & 1.14 & $-$1.79 & 0.52 & 0.20 & 0.32 & & $-$0.66 & $-$0.72 & 0.06 \\
 HD 166161    & 5350 & 2.56 & $-$1.23 & 1.19 & 0.84 & 0.35 & & $-$0.41 & $-$0.48 & 0.07 \\
 HD 171496    & 4950 & 2.37 & $-$0.67 & 1.60 & 1.41 & 0.19 & & 0.17 & 0.11 & 0.06 \\
 HD 175305    & 4770 & 1.80 & $-$1.73 & 0.06 & $-$0.28 & 0.34 & & $-$0.82 & $-$0.89 & 0.07 \\
 HD 187111    & 4270 & 1.05 & $-$1.97 & 0.21 & $-$0.10 & 0.31 & & $-$0.82 & $-$0.88 & 0.06 \\
 HD 201891    & 5910 & 4.19 & $-$1.09 & 1.58 & 1.25 & 0.33 & & $-$0.17 & $-$0.22 & 0.05 \\
 HD 204543    & 4670 & 1.49 & $-$1.87 & 0.41 & 0.05 & 0.36 & & $-$0.98 & $-$1.05 & 0.07 \\
 HD 206739    & 4650 & 1.78 & $-$1.72 & 0.69 & 0.38 & 0.31 & & $-$0.55 & $-$0.62 & 0.07 \\
 HD 210295    & 4750 & 2.50 & $-$1.46 & 1.04 & 0.72 & 0.32 & & $-$0.27 & $-$0.34 & 0.07 \\
 HD 214925    & 4050 & 0.00 & $-$2.08 & $-$0.13 & $-$0.50 & 0.37 & & $-$0.97 & $-$1.09 & 0.12 \\
 HD 220838    & 4300 & 0.60 & $-$1.80 & 0.39 & 0.05 & 0.34 & & $-$0.86 & $-$0.93 & 0.07 \\
 HD 221170    & 4510 & 1.00 & $-$2.16 & 0.21 & $-$0.09 & 0.30 & & $-$0.80 & $-$0.86 & 0.06 \\
 HD 235766    & 4650 & 1.20 & $-$1.93 & 0.42 & 0.10 & 0.32 & & $-$0.80 & $-$0.86 & 0.06 \\
 HE 1523-0901$^3$ & 4630 & 1.00 & $-$2.95 & 0.11 & $-$0.34$^4$& 0.45 & & $-$0.57 & $-$0.62 & 0.05 \\
\noalign{\smallskip}\hline \noalign{\smallskip}
\multicolumn{11}{l}{ \ $^1$ from the calculations with \kH\,= 0.1,} \\ 
\multicolumn{11}{l}{ \ $^2$ stellar parameters and LTE abundances from \citet{Sitnova2011},} \\
\multicolumn{11}{l}{ \ $^3$ stellar parameters and europium LTE abundance from \citet{he1523},} \\
\multicolumn{11}{l}{ \ $^4$ Frebel et al. (2012, in preparation).} \\
\end{tabular}
\end{table*}

\section{Constraints on the Pb production mechanisms in the early Galaxy}\label{sec:pb_stars}

In this section, we revise the Pb abundances of metal-poor stars
available in the literature and investigate the metallicity behaviour
of Pb/Eu abundance ratio. This ratio is particularly sensitive to
whether the production of lead occurred in a pure r-process or whether
there was a non-negligible s-process contribution. The r-nuclei of
lead are produced through several decay channels. At the time when the
r-process event stops (likely after a few seconds), there is the
direct $\beta^-$ decay of very neutron-rich isobaric nuclei with
$A=206-208$ to $^{206}{\rm Pb}$, $^{207}{\rm Pb}$, and $^{208}{\rm
  Pb}$. Then there is the $\alpha$- and $\beta$-decay of nuclei with
$210 \le A \le 231$ and $A = 234$ back to Pb, which operates shortly
after the termination of the r-process, within few million years. And
finally the radioactive decay of the long-lived Th and U isotopes
produced Pb over the course of almost the age of the Galaxy.  As
stressed by \citet{Roederer2009}, no Pb isotopes can be measured
individually due to the relatively small isotope shifts ($\le
0.02-0.03$\,\AA) and overall weakness of the \ion{Pb}{i} lines.

Hence, for the total elemental abundances, the classical waiting-point
(WP) r-process model of \citet{Kratz2007} predicts $\eps{r,0}({\rm
  Pb/Eu}) = 0.61$ in a newly born star and $\eps{r,13}({\rm Pb/Eu}) =
0.70$ in a 13~Gyr old star \citep{Roederer2009}. Another estimate of a
pure r-process production ratio can be obtained using the Solar
r-residuals. With an s-process contribution to the solar Pb of
81\,\%\ \citep{2001ApJ...549..346T}\footnote{These authors actually
  indicated 91\,\%. However, using s-fractions of 90, 60, 77, and
  89\,\%\ for the $^{204}{\rm Pb}$, $^{206}{\rm Pb}$, $^{207}{\rm
    Pb}$, and $^{208}{\rm Pb}$ isotopes, respectively, from their
  Table~3, we calculated an s-fraction of 81\,\%\ for the total lead
  abundance.} and to the solar Eu of 6\,\%\ \citep{Travaglio1999}, we
computed the Solar System r-process ratio $\eps{SSr}({\rm Pb/Eu}) =
0.82$. When computing s-fractions,
\citet{Travaglio1999,2001ApJ...549..346T} used the meteoritic isotope
abundances from \citet{AG1989}.  For comparison, the ratio of the
meteoritic total abundances is $\eps{met}({\rm Pb/Eu}) = 1.51$ and
1.53 according to \citet{AG1989} and \citet{Lodders2009},
respectively.

The largest data set on stellar Pb abundances was compiled by
\citet{Roederer2010}. Their data are based on the analysis of
\ion{Pb}{i} 4057\,\AA. 
Before correcting these abundances, we excluded all stars with only upper
limits for Pb abundance, the carbon-enhanced stars with [C/Fe] $>
0.3$, and also VMP stars enriched in s-process material, with
$\eps{}$(Pb/Eu)$\ge +1.8$ at [Fe/H] $< -2$ (see small blue
``$\times$'' in their Fig.~3) from the \citet{Roederer2010}
sample. Regarding obtaining carbon abundances for these stars to check
upon, we also relied on the \citet{Simmerer2004} study, which was used
by \citet{Roederer2010} to compile most of their stellar sample.  In
total, 49 stars were selected in the $-2.26 \le$ [Fe/H] $< -0.59$
metallicity range. We added to them the two r-II stars, i.e.,
CS~31082-001 \citep{hill2002,plez_lead} and HE~1523-0901
\citep[][2012, in preparation]{he1523}, and also the
halo star HD~29907 from \citet{Sitnova2011}. The stars used are listed
in Table~\ref{pb_corr} with stellar parameters and LTE element
abundances taken from the cited papers and with non-LTE abundance
corrections obtained in this study.  The non-LTE calculations were
performed with the MARCS model atmospheres
\citep{Gustafssonetal:2008}. Where necessary, the MARCS model
structures were interpolated for given $\Teff$, $\log g$, and [Fe/H]
using a FORTRAN-based routine written by Thomas Masseron ({\tt
  http://marcs.astro.uu.se/software.php}). For each star, the SE
calculations were performed using its given LTE element abundance.

For self-consistency, we also revised the europium abundances of the
stars in our sample.  The non-LTE calculations for \ion{Eu}{ii} were
performed using the method treated by \citet{mash_eu}. Everywhere,
\kH\ = 0.1 was adopted. In the stellar parameter range with which we
are concerned, non-LTE leads to weakened \ion{Eu}{ii} lines and
positive abundance corrections. We computed $\Delta_{\rm NLTE}$ for
the seven lines,
i.e., \ion{Eu}{ii} 3724, 3819, 4129, and 4205\,\AA\ with \Eexc\ = 0
and \ion{Eu}{ii} 3907, 4435, and 4522\,\AA\ with \Eexc\ =
0.21\,eV. Table~\ref{pb_eu_corr} presents the resulting values for \ion{Eu}{ii} 4129\,\AA\ in the grid of model atmospheres
 characteristic of the Galactic halo stars. Table~\ref{pb_corr} does the same for the
stars in our sample.
For each line, $\Delta_{\rm NLTE}$ grows toward lower
metallicity and lower surface gravity, although, it is overall small
and does not exceed 0.17~dex for different lines. For example,
$\Delta_{\rm NLTE}$ ranges between 0.02 and 0.06~dex for different
lines in the 5600/3.75/$-0.59$ model, between 0.04 and 0.14~dex in the
4825/1.5/$-2.9$ model, and between 0.05 and 0.17~dex in the
4050/0.0/$-2.08$ model. 
 For each model atmosphere, $\Delta_{\rm NLTE}$
of \ion{Eu}{ii} 4129\,\AA\ turned out to well represent the mean
non-LTE correction from all the computed lines. For a given star, we
used it as the mean non-LTE correction and applied it to the observed
europium LTE abundance. 

\begin{figure}
\flushleft 
\hspace{-5mm}
  \resizebox{88mm}{!}{\includegraphics{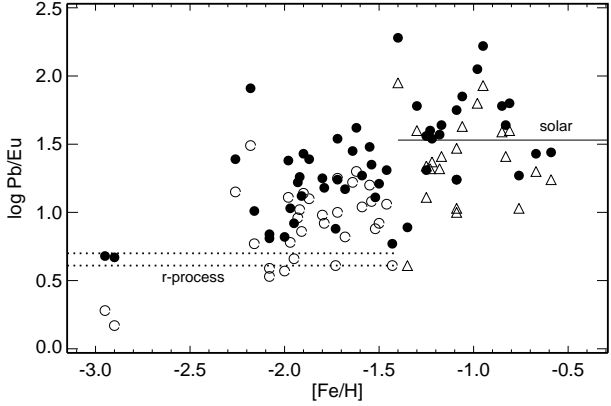}}
\caption{\label{fig:stellar_pb} Stellar non-LTE (filled circles) and
  LTE (open circles for the stars with $\log g \le 3$ and triangles
  for $\log g > 3$) Pb/Eu abundance ratios as a function of
  metallicity. The solid line indicates the Solar system meteoritic
  value. The two dotted lines correspond to a pure r-process
  production of Pb and Eu predicted by the waiting-point r-process
  model for a newly born (the low-lying curve) and 13~Gyr old (the
  up-lying curve) star as given by \citet{Roederer2009}. The non-LTE
  abundances were calculated with \kH\ = 0.1.  }
\end{figure}

The non-LTE and LTE Pb/Eu abundance ratios are plotted in
Fig.\,\ref{fig:stellar_pb}. As shown by \citet{hill2002} for
CS~31082-001 and \citet{he1523} for HE~1523-0901, the heavy elements
beyond the iron group in these stars originate from the r-process, and
the observed element abundance ratios can be used to test theoretical
nucleosynthesis models. We found that both stars have very similar
Pb/Eu abundance ratios at $\eps{NLTE}$(Pb/Eu) = 0.67 and 0.68,
respectively, and that the WP r-process model for a 13~Gyr old star
\citep{Kratz2007,Roederer2009} perfectly reproduces the
observations. We also stress that the Pb/Eu abundance ratios of the
r-II stars match the Solar r-process ratio $\eps{SSr}({\rm Pb/Eu}) =
0.82$ within the error bars. According to \citet{plez_lead}, the
uncertainty in the Pb abundance of CS~31082-001 is 0.15~dex. This
provides strong evidence for universal Eu and Pb relative r-process
yields during the Galactic history. 

Despite the large scatter in the abundances, a clear upward trend is
seen in the $\eps{}({\rm Pb/Eu})$ - [Fe/H] plane for all stars with
[Fe/H] $> -2.3$. The Pb/Eu abundance ratio grows from $\eps{NLTE}({\rm
  Pb/Eu}) \simeq 0.8$ at [Fe/H] $= -2$ up to the solar value at [Fe/H]
$= -0.6$.  Following \citet{Roederer2010}, we separated these stars in
two groups depending on their metallicity. The low-metallicity sample
(hereafter, LMS) includes 29 stars in the $-2.26 \le$ [Fe/H] $< -1.40$
metallicity range, and the mildly metal-poor sample ($-1.4 \le$ [Fe/H]
$\le -0.59$, hereafter, MMPS) includes 21 stars. It is worth noting
that each subsample is highly surface gravity biased due to a common
selection effect.  Namely, most LMS stars are cool giants, while the
MMPS stars are mainly dwarf stars.

As seen in Fig.\,\ref{fig:stellar_pb}, the LMS stars have
significantly higher Pb/Eu ratios compared to that of the r-II stars,
independent of either LTE or non-LTE. Excluding a clear outlier, i.e.,
the star HD\,108317 ([Fe/H] = $-2.18$, $\eps{NLTE}({\rm Pb/Eu}) =
1.91$), we calculated the mean $\eps{LTE}({\rm Pb/Eu}) = 0.93\pm0.23$
and $\eps{NLTE}({\rm Pb/Eu}) = 1.19\pm0.24$ from the 28 LMS stars. We
shall conclude that the s-process in AGB stars started to produce lead
before the Galactic metallicity grew to [Fe/H] $= -2.3$. The
low-metallicity s-process nuclesynthesis calculations
\citep{Roederer2010,2011arXiv1112.2757L} do not contradict this
observational finding. \citet{Roederer2010} presented the surface
composition at the end of the AGB phase for a number of
intermediate-mass models with [Fe/H] $= -2.3$ and $M = 4.5-6
M_\odot$. The minimum value produced by those models is $\eps{}({\rm
  Pb/Eu}) = 1.9$. Given sufficient time, their contributions could
result in raising the Galactic Pb/Eu abundance ratio above the pure
r-process one.  Earlier, from analyses of the stellar (Ba, La,
Nd)/(Eu, Dy) abundance ratios, \citet{2000ApJ...544..302B} and
\citet{Simmerer2004} obtained evidences for that the s-process might
be active as early as [Fe/H] $= -2.3$ and $-2.6$, respectively.

For the MMPS stars, the obtained mean, $\eps{NLTE}({\rm Pb/Eu}) =
1.58\pm0.31$, is very close to the Solar System value, $\eps{met}({\rm
  Pb/Eu}) = 1.53$ \citep{Lodders2009}. This supports the theoretical
result of \citet{2001ApJ...549..346T} who predicted that the AGB stars
with [Fe/H] $\simeq -1$ made the greatest contribution to the solar
abundance of s-nuclei of lead. It is worth noting that LTE leads to
0.22~dex lower Pb/Eu abundance ratio for the MMPS stars. The star
G\,58-25, with [Fe/H] = $-1.40$ and $\eps{NLTE}({\rm Pb/Eu}) = 2.28$,
was not included in the mean calculations.

Generally, our findings are stable with respect to a variation in the stellar parameters that correspond to reasonable observational uncertainties and a variation in the \kH\ value. For example, for CS~31082-001, a change of +100~K in $\Teff$ (a $\sim2 \sigma$ error in \citet{hill2002}) produces a change in the calculated Pb non-LTE abundance of +0.16~dex. \ion{Pb}{i} 4057\,\AA\ is insensitive to a variation in $\logg$ and microturbulence velocity, $\Vmic$. The total effect of varying $\Teff$, $\logg$ (by +0.3~dex), and $\Vmic$ (by +0.2\,\kms) on the derived Eu abundance was estimated by \citet{hill2002} as to be +0.12~dex. Thus, the Pb/Eu abundance ratio is only slightly affected by the uncertainties in stellar parameters. The non-LTE calculations performed for
CS~31082-001 with \kH\ = 0 and 1 result in only 
0.04 and 0.02~dex lower $\eps{}$(Pb/Eu) values compared to that for
\kH\ = 0.1. To evaluate the uncertainty in the Pb/Eu abundance ratios of the LMS and MMPS stars, we chosen the two representative stars HD\,3008 (4250/0.25/$-2.08$) and G\,102-020 (5250/4.44/$-1.25$). \citet{Simmerer2004} estimated that the stellar parameters of their sample are internally consistent to $\Delta\Teff = 100$~K, $\Delta\logg = 0.25$, and $\Delta\Vmic = 0.1$\,\kms. An upward revision of the stellar parameters for HD\,3008 results in a cumulative change of +0.11~dex in $\eps{Pb}$ and +0.10~dex in $\eps{Eu}$. For G\,102-020, the abundance errors caused by the uncertainties in the stellar parameters also have a common sign for Pb and Eu and amount to 0.09 and 0.06~dex, respectively. 

We note that we have not considered effects on the abundances as a
result of using 3D hydrodynamical model atmospheres. However, those
effects were estimated for \ion{Pb}{i} 4057\,\AA\ (\Eexc\ = 1.32\,eV)
and the \ion{Eu}{ii} lines arising from the ground state using the
3D-1D LTE abundance corrections calculated by \citet{Collet2007} and
\citet{Hayek2011} for a sample of ``fictitious'' atomic (\ion{Na}{i},
\ion{Mg}{i}, \ion{Ca}{i}, \ion{Fe}{i}, and \ion{Fe}{ii}) lines at
selected wavelengths, with varying lower-level excitation potentials
and line strengths. Overall, 3D-1D corrections are negative for the
lines of neutral atoms and decrease in absolute value with increasing
\Eexc. In the 4717/2.2/$-1$ models, they have similar values for all
the chemical species, i.e., 3D-1D varies between $-0.20$ and
$-0.28$~dex for the lines with \Eexc\ = 0 and between $-0.10$ and
$-0.18$~dex for \Eexc\ = 2\,eV. The 3D effects grow toward lower
metallicity, however, with different rate for different atoms. In the
4732/2.2/$-2$ and 4858/2.2/$-3$ models, the 3D-1D corrections are
similar for \ion{Mg}{i} and \ion{Fe}{i}, but different from that for
\ion{Na}{i} and \ion{Ca}{i}. For the weak \ion{Na}{i} line with
\Eexc\ = 1.3\,eV and an equivalent width of EW $< 20$~m\AA, 3D-1D
$\simeq -0.12$~dex in the $-3 \le$ [Fe/H] $\le -1$ metallicity
range. For the weak \ion{Fe}{i} line with the same excitation
potential, 3D-1D $\simeq -0.4, -0.3$, and $-0.17$~dex in the [Fe/H] $=
-3, -2$, and $-1$ model, respectively. The 3D effects are overall weak
for the \ion{Fe}{ii} lines arising from the ground state. For example,
for the moderate strength line (EW $\simeq 50$~m\AA), 3D-1D $= -0.03,
+0.15$, and $+0.15$~dex in the [Fe/H] $= -3, -2$, and $-1$ model,
respectively. We assume that 3D-1D corrections for the investigated
\ion{Eu}{ii} lines are similar to that for the ``fictitious''
\ion{Fe}{ii} line with \Eexc\ = 0\,eV. The 3D-1D correction for
\ion{Pb}{i} 4057\,\AA\ lies, probably, between those for \ion{Na}{i}
and \ion{Fe}{i}. In case it is as large as that for \ion{Na}{i},
accounting for the 3D effects results in approximately 0.1~dex lower
Pb/Eu abundance ratio for the two r-II stars and 0.25~dex lower Pb/Eu
value for the low-metallicity and mildly metal-poor samples. With such
changes, our conclusions remain valid. If the 3D-1D correction for
\ion{Pb}{i} 4057\,\AA\ is as large as that for \ion{Fe}{i}, the mean
Pb/Eu abundance ratios of the r-II stars and the LMS stars are
expected to be approximately 0.4~dex lower compared with the
corresponding NLTE+1D values.
However, in this case, it would be hard to understand why the Pb/Eu
abundance ratio of the strongly r-process enhanced stars is
significantly lower (by 0.5~dex) compared with the Solar System
r-process value.

We note here that full 3D non-LTE computations have so far only been
performed for \ion{Li}{i} in the Sun and metal-poor stars
\citep{Asplund2003_Li,Barklem2003_Li,Cayrel2007_Li,Sbordone2010_Li}
and for \ion{O}{i} in the Sun \citep{Asplund2004}. It was found that
the 3D+non-LTE effects are overall small for the \ion{Li}{i} resonance
line because the line strengthening due to the cooler temperatures in
the upper atmospheric layers of the 3D model on the one hand, and line
weakening from increased overionization on the other hand, largely
cancel each other out. For \ion{O}{i}, the non-LTE abundance
corrections turned out similar in the 3D and 1D solar model
atmospheres. In contrast, \citet{Shchukina2005} predicted stronger
non-LTE effects for \ion{Fe}{i} in 3D than in the 1D model, based on
their 1.5D+non-LTE calculations for the Sun and HD\,140283
(5700/3.7/$-2.5$). Full 3D non-LTE computations need to be done for
\ion{Pb}{i} to decide what is a difference in Pb/Eu abundance ratio
between 3D+non-LTE and 1D+non-LTE cases and how this might affect our
conclusions.

\section{Non-LTE calculations for \ion{Th}{ii}}\label{sec:nlte_th}

\subsection{Thorium model atom}

\begin{figure*}
\resizebox{170mm}{!}{\includegraphics{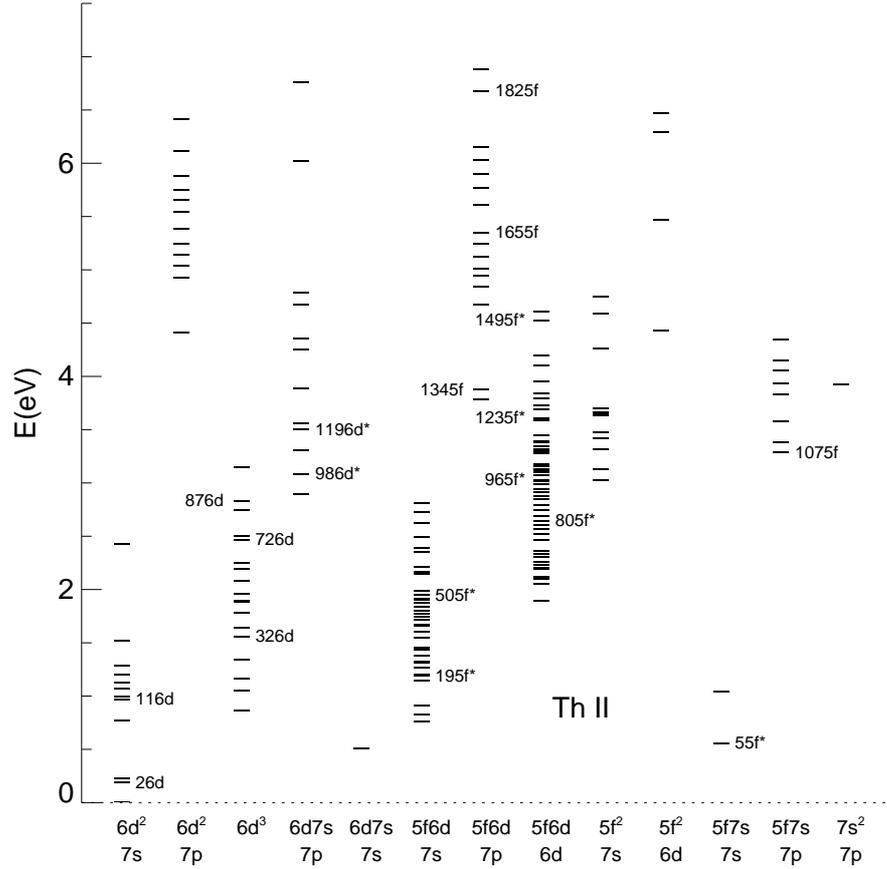}}
\caption{\label{Th_atom} Model atom of \ion{Th}{ii}. The names of the
  levels important for understanding the non-LTE mechanisms are
  indicated.}
\end{figure*}

{\it Energy levels.} In the range of stellar parameters that we are
considering here, the majority of the element exists as
\ion{Th}{ii}. For example, the fraction of neutral thorium, with its
ionization energy $\chi_{\rm ion}$ = 6.3~eV, does not exceed 10$^{-3}$
throughout the 4500/1/$-3$ atmosphere and $6\cdot$10$^{-3}$ throughout
the solar atmosphere. We thus neglected \ion{Th}{i} when constructing
the model atom.  The \ion{Th}{ii} term structure is produced by
multiple electronic configurations (see Fig.~\ref{Th_atom}) and
consists of thousands energy levels. The ground configuration is
6d$^2$7s. We used the energy levels with \Eexc\ $\le$ 6.88~eV, which
amount to 416 levels from \citet{Blaise1992}.
The higher excitation levels play a minor role in the population and
depopulation of \ion{Th}{ii}, because the next ionization stage,
\ion{Th}{iii}, represents a minor fraction of the thorium
abundance. For example, the contribution of \ion{Th}{iii} to thorium
abundance is smaller than 0.2 / 1.5~\%\ in the cool VMP / solar models
outside $\log\tau_{5000} = 0$.  The levels with common parity and
close energies were combined whenever the energy separation is smaller
than 0.01~eV at \Eexc\ $<$ 4~eV and smaller than 0.1~eV at higher
\Eexc. The final model atom includes 184 levels of \ion{Th}{ii}
(Fig.~\ref{Th_atom}).

The \ion{Th}{iii} ground state 5f6d$^3$H$^\circ_4$ is a poor
representation of the \ion{Th}{iii} ionization stage. The odd 5f6d and
5f7s and the even 6d$^2$ and 6d7s configurations of \ion{Th}{iii}
produce many low-excitation levels. The energy of 6d$^2$ $^3{\rm
  F}_2$, \Eexc\ = 63.2~cm$^{-1}$, is the smallest distance between the
lowest levels of both parities ever found in an atomic spectrum. As a
result, the \ion{Th}{iii} partition function ($U$) is significantly
larger compared with the statistical weight $g = 9$ of the ground
state. For example, $U$(\ion{Th}{iii}) = 49.14 at the temperature $T
=$ 4990~K and grows to $U$(\ion{Th}{iii}) = 81.11 at $T =$
8318~K. These values were computed
using the \ion{Th}{iii} energy levels from
\citet{Blaise1992}. Assuming that the low-excitation levels of
\ion{Th}{iii} are thermally coupled to the ground state, we included
\ion{Th}{iii} in the model atom as a single state with $g = 49$, while
keeping the energy of the ground state.

{\it Radiative rates.}  Accurate $gf-$values based on measured natural
radiative lifetimes and branching ratios were obtained by
\citet{Nilsson_Th} for 180 transitions between the lowest (\Eexc\ = 0
-- 2.25~eV) even parity levels of the 6d$^2$7s, 6d7s$^2$, and 6d$^3$
electronic configurations and the intermediate energy (\Eexc\ = 2.26
-- 4.08~eV) odd parity levels. For another 1244 transitions, we used
$gf-$values of \citet{Kur94b} which are accessible via the {\sc VALD}
database \citep{vald}. These two sources provide the data for 955
$b-b$ transitions between combined levels in our model atom. They all
were included in SE calculations. We refer to this atomic model as
Th-1.

The second atomic model, Th-2, includes the extended system of $b-b$
radiative transitions. We estimated the total number of allowed
transitions in the system of 416 \ion{Th}{ii} levels assuming that
every transition between the different parity levels with $\Delta J =
0, \pm 1$ is an allowed one. The exception are the $J = 0
\leftrightarrow J = 0$ transitions. This gives 20\,214 transitions,
and for 18\,790 of them, oscillator strengths ($f_{ij}$) are
missing. The 1\,562 of the latter transitions, with the energy
separation of less than 0.25~eV, were treated as forbidden ones
because of insignificant contribution of the radiative to the total
transition rate, irrespective of the transition $gf-$value. For every
of the remaining 17\,228 transitions, $f_{ij} = 10^{-4}$ was
adopted. Such a choice was based on the inspection of the known
$gf-$values for \ion{Th}{ii}. No transition from the
\citet{Nilsson_Th} and \citet{Kur94b} data sets has $f_{ij} \le
10^{-4}$. In this atomic model, the 5\,661 $b-b$ transitions were
included in SE calculations.

The photoionization cross-sections were computed using the hydrogen
approximation formula with n = n$_{\rm eff}$, because no accurate data is
available for the \ion{Th}{ii} levels.

{\it Collisional rates.} The same formulas as for \ion{Pb}{i} were
employed for the transitions in \ion{Th}{ii}, with $\Upsilon$ = 2 for
the forbidden ones.

\subsection{Departures from LTE for \ion{Th}{ii}}

\begin{figure}
\flushleft 
\resizebox{88mm}{!}{\includegraphics{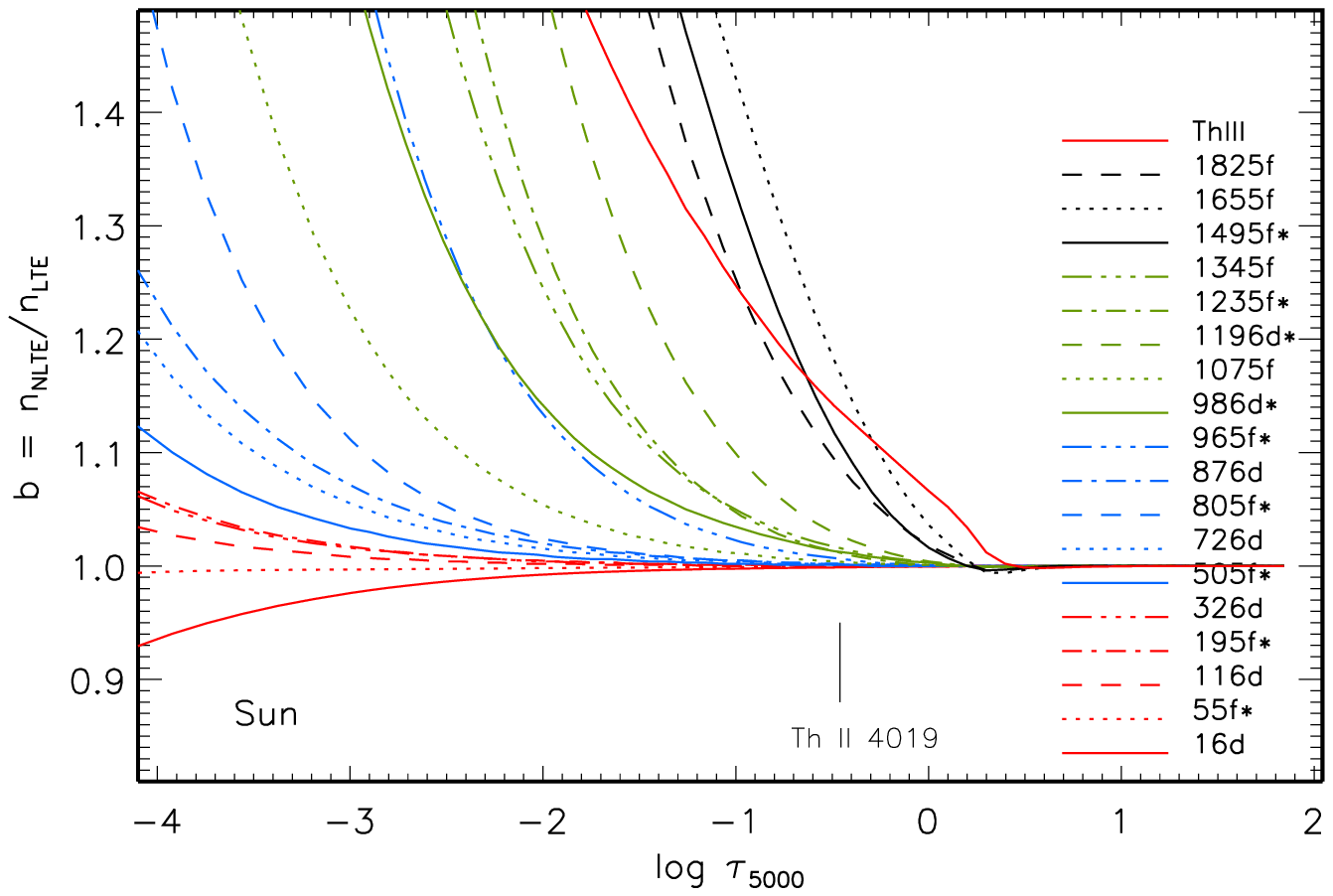}}
\resizebox{88mm}{!}{\includegraphics{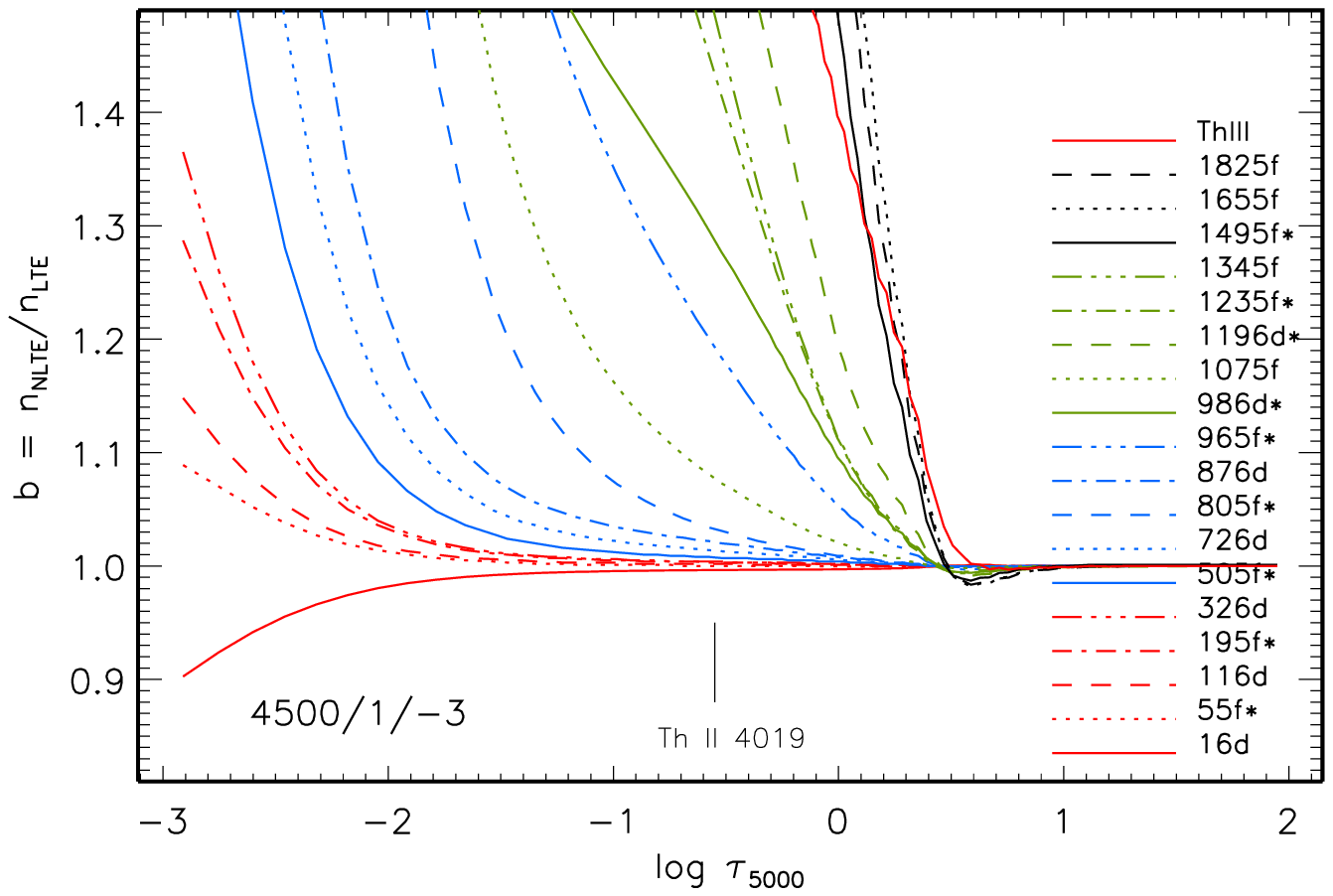}}
\caption[]{The same as in Fig.\,\ref{Fig:bf_pb} for the selected
  levels of \ion{Th}{ii} and the \ion{Th}{iii} ground state from the
  calculations with the Th-2 atomic model. The vertical lines indicate
  the locations of line core formation depths for \ion{Th}{ii}
  4019.} \label{Fig:bf_th}
\end{figure}

The non-LTE calculations for \ion{Th}{ii} were performed with the MARCS 
solar model atmosphere and with the models representing the
atmospheres of VMP stars with [Fe/H] $= -2$ and $-3$ (see
Table\,\ref{th_corr}).  Figure\,\ref{Fig:bf_th} shows the departure
coefficients of the selected \ion{Th}{ii} levels in the 5780/4.44/0
and 4500/1.0/$-3$ models. We found that the lowest 72 levels, i.e., up
to the level denoted as 726d (\Eexc\ = 2.46~eV), keep nearly TE
populations ($b \simeq 1$) in the atmosphere inside $\log\tau_{5000} =
-2$ and $-1$ for the solar and VMP models, respectively. For all the
higher excitation levels, $b > 1$. This is because the population of
the odd parity 5f6d$^2$ and 6d7s7p levels with \Eexc\,= 3 - 4.8~eV is
controlled by the strong radiative transitions from the three lowest
6d$^2$7s states and the 6d7s$^2$ level, such as 6d$^2$7s$^4$F$_{3/2}$
- 6d7s7p (\Eexc\,= 3.084~eV, $J = 5/2$) denoted in our model atom as
16d - 986d* and producing the line at 4019~\AA, 16d - 965f*
(4094~\AA), 26d - 1196d* (3741~\AA), and 36d - 1495f*
(2895~\AA). These transitions are pumped by the ultraviolet $J_\nu -
B_\nu(T_e)$ excess radiation and produce enhanced excitation of the
upper levels far inside the atmosphere. This overpopulation is
redistributed to the even parity high excitation levels through
inelastic collisions.

The Th-2 atomic model includes more pumping transitions compared to
the Th-1 one, and the departures from LTE are expected to be stronger
when using the Th-2 model.  We checked the populations of the lower
and upper levels of the \ion{Th}{ii} 4019~\AA\ transition. In the line
formation layers, around $\log\tau_{5000} = -0.45$ and $-0.55$ in the
solar and 4500/1.0/$-3$ models, respectively,
the difference in non-LTE populations between the Th-2 and Th-1 atomic
models is negligible for the \ion{Th}{ii} ground state. This is easy
to understand because the lowest \ion{Th}{ii} levels contain the
majority of the element, and no mechanism is able to significantly
change their populations compared to the TE ones. For the upper level,
986d*, a small difference of 3\,\%\ between using Th-2 and Th-1 was
found in the VMP model, and very similar populations were obtained in
the solar model.

The \ion{Th}{ii} lines used in abundance analyses all arise from
either the ground state or low-excitation levels with \Eexc\ $\le$
0.51~eV, for which $b \simeq 1$ holds in the line formation layers
(Fig.\,\ref{Fig:bf_th}). Nevertheless, for each line, non-LTE leads to
its weakening compared to the LTE strength, owing to the
overpopulation of the upper level relative to the TE population that
results in $b_u/b_l > 1$ and the rise of the line source function
above the Planck function in the line-formation layers. Overall, the
obtained non-LTE abundance corrections are positive.  For the Sun, the
departures from LTE are small irrespective of the applied \kH\ value,
such that $\Delta_{\rm NLTE}$(\ion{Th}{ii} 4019~\AA) = 0.06 and
0.00~dex in case of \kH\ = 0 and 1, respectively. In the VMP models,
the non-LTE correction is very sensitive to treatment of inelastic
collisions with \ion{H}{i} atoms. For example, for 4500/1/$-3$, it
amounts to $\Delta_{\rm NLTE} = +0.52$\,dex in the calculations with
pure electronic collisions and it varies between $\Delta_{\rm NLTE} =
+0.12$ and $+0.03$\,dex when collisions with \ion{H}{i} atoms are
taken into account with \kH\ = 0.1 to 1. The use of the Th-1 atomic
model results in 0.01~dex smaller non-LTE corrections.

\begin{table}
 \centering
 \caption{\label{th_corr} Non-LTE abundance corrections (dex) for the \ion{Th}{ii} lines from the calculations with \kH\,= 0.1.}
  \begin{tabular}{cccccc}
   \hline\noalign{\smallskip}
$\Teff$ & $\log g$ & [Fe/H]  & [Th/Fe]$^1$ & \multicolumn{2}{c}{$\Delta_{\rm NLTE}$} \\
\noalign{\smallskip} \cline{5-6} \noalign{\smallskip}
 & &    &         & 4019\,{\AA} & 4086\,{\AA} \\
   \hline\noalign{\smallskip}
5780 & 4.44 & ~0  & 0.09$^2$ & 0.01 & -$^3$ \\
4000 & 0.0 & $-2$ & 0.4~  & 0.21 & 0.14 \\ 
4000 & 0.5 & $-2$ & 0.4~  & 0.13 & 0.08 \\
4250 & 0.5 & $-2$ & 0.4~  & 0.16 & 0.09 \\
4250 & 1.0 & $-2$ & 0.4~  & 0.10 & 0.06 \\
4500 & 1.0 & $-2$ & 0.4~  & 0.12 & 0.07 \\
4500 & 1.5 & $-2$ & 0.4~  & 0.07 & 0.04 \\
4750 & 1.5 & $-2$ & 0.4~  & 0.09 & 0.05 \\
4750 & 2.0 & $-2$ & 0.4~  & 0.06 & 0.03 \\
5250 & 2.5 & $-2$ & 0.4~  & 0.05 & - \\
4500 & 1.0 & $-3$ & 1.75 & 0.12 & 0.06 \\
4500 & 1.5 & $-3$ & 1.75 & 0.07 & 0.03 \\
4750 & 1.0 & $-3$ & 1.75 & 0.15 & 0.09 \\ 
4750 & 1.5 & $-3$ & 1.75 & 0.09 & 0.05 \\
\noalign{\smallskip} \hline \noalign{\smallskip}
\multicolumn{6}{l}{ \ $^1$ Thorium abundance used in non-LTE calculations} \\
\multicolumn{6}{l}{ \ $^2$ Absolute thorium abundance, $\eps{Th}$} \\
\multicolumn{6}{l}{ \ $^3$ Calculated equivalent width is smaller than 1 m\AA} \\ 
\end{tabular}
\end{table}

Table~\ref{th_corr} presents $\Delta_{\rm NLTE}$ for the two lines,
\ion{Th}{ii} 4019 and 4086~\AA, from the calculations with the Th-2
atomic model and \kH\ = 0.1. For a given $\Teff$, the departures from
LTE grow toward lower surface gravity due to decreasing contribution
of the collisional to the total transition rates. Maximum $\Delta_{\rm
  NLTE}$ were calculated for the 4000/0.0/$-2$ model, and they amount
to +0.21~dex for \ion{Th}{ii} 4019\,{\AA} and +0.14~dex for
\ion{Th}{ii} 4086\,{\AA}. The metallicity effect is less pronounced in
Table~\ref{th_corr} because, for the [Fe/H] $=-3$ models, the non-LTE
calculations were performed with a higher Th abundance compared with
that in the [Fe/H] $=-2$ models.

\section{Conclusions}\label{Conclusions}

We built a comprehensive model atom for neutral lead using atomic data
for the energy levels and transition probabilities from laboratory
measurements and theoretical predictions. The non-LTE calculations for
\ion{Pb}{i} were performed for the first time for the Sun and for a
set of stellar parameters characteristic of metal-poor stars in the
$-2.95 \le$ [Fe/H] $\le -0.59$ metallicity range. We found that the
main non-LTE mechanism for \ion{Pb}{i} is the ultraviolet
overionization, although radiative pumping of the strong transitions
arising from the low-lying levels tends to produce enhanced excitation
of the \ion{Pb}{i} levels with \Eexc\ $>$ 4~eV. Overall, non-LTE leads
to weakened \ion{Pb}{i} lines and positive non-LTE abundance
corrections. The departures from LTE grow with decreasing metallicity
and, for a given metallicity, increase toward higher effective
temperatures and lower surface gravities. The main source of the
uncertainty in the calculated non-LTE abundance corrections are poorly
known inelastic collisions with \ion{H}{i} atoms. From analysis of
either solar or stellar \ion{Pb}{i} lines we can not empirically
constrain their efficiency in SE calculations. With \kH\,= 0.1, chosen
based on our estimates for \ion{Ca}{i}-\ion{Ca}{ii} and
\ion{Fe}{i}-\ion{Fe}{ii} \citep{mash_ca,mash_fe}, $\Delta_{\rm NLTE}$
of \ion{Pb}{i} 4057\,\AA\ ranges between 0.16~dex in the solar model
and 0.56~dex in the 4825/1.5/$-2.9$ model.

We also built a model atom for singly-ionized thorium using atomic
data for the energy levels 
from laboratory measurements. The non-LTE calculations for \ion{Th}{ii}
were performed for the first time for the Sun and for the small grid
of model atmospheres with [Fe/H] $= -2$ and $-3$. In contrast to
\ion{Pb}{i}, \ion{Th}{ii} is the majority species in the stellar
parameter range that we have covered, and the main non-LTE mechanism
for \ion{Th}{ii} is connected to the pumping transitions arising from
the low-excitation levels, with \Eexc\ $<$ 1~eV. Overall, non-LTE
leads to weakened \ion{Th}{ii} lines and positive non-LTE abundance
corrections. Compared with \ion{Pb}{i}, the departures from LTE for
\ion{Th}{ii} are even more sensitive to the efficiency of collisions
with \ion{H}{i} atoms.
With \kH\ = 0.1, $\Delta_{\rm NLTE}$ of the investigated \ion{Th}{ii}
lines nowhere exceeds 0.21~dex.

We caution against using the LTE assumption in stellar analyses of Pb
abundance and also the elemental abundance ratios involving Pb and
some other element observed in the lines of their majority species,
such as Eu, Th, or U.

Taking advantage of our SE approach for \ion{Pb}{i}, we obtained good
agreement, within 0.03\,dex, between meteoritic and absolute solar
abundances from the calculations with the \citet{HM74} solar model
atmosphere and \kH\,= 0.1. Our final value is $\eps{Pb,\odot}$ = 2.09.

We then calculated the non-LTE abundances of Pb and Eu for the
\citet{Roederer2010} metal-poor stars and complemented that sample
with HE~1523-0901 \citep{he1523} and HD~29907
\citep{Sitnova2011}. We found that the two strongly r-process enhanced
(r-II) stars have very similar Pb/Eu abundance ratios, with the mean
$\eps{NLTE}$(Pb/Eu) = 0.68$\pm$0.01, and the waiting-point r-process
model as presented by \citet{Roederer2009} reproduces the observations
very well. The revised Pb/Eu abundance ratios of the r-II stars match,
within the error bars, the corresponding solar r-process ratio.  Thus,
{\it the universality of the r-process was proved} not only for the
second r-process peak elements from Ba to Hf as found earlier by
\citet[][and references therein]{sneden1996,hill2002,Sneden2008}, but
also for the heavier element Pb.

It was found that the stars in the $-2.3 \le$ [Fe/H] $< -0.6$
metallicity range, reveal a clear upward trend in the $\eps{}({\rm
  Pb/Eu})$ - [Fe/H] plane. Following \citet{Roederer2010}, we
separated these stars in two groups depending on their
metallicity. The lower metallicity stars ([Fe/H] $< -1.4$) turned out
to have, on average, 0.51~dex higher Pb/Eu ratios compared with that
of the r-II stars. This led us to conclude that {\it the s-process
  production of lead started before the Galactic metallicity grew to
  ${\rm [Fe/H]} = -2.3$}. It is worth noting that the studies of
\citet{2000ApJ...544..302B} and \citet{Simmerer2004} who obtained
(Ba, La, Nd)/(Eu, Dy) abundance ratios of metal-poor
stars, also found that the s-process has been active as early as
[Fe/H] $= -2.3$ and $-2.6$, respectively. 

The mean, $\eps{NLTE}({\rm Pb/Eu}) = 1.58\pm0.31$, for the mildly
metal-poor subsample ($-1.4 \le$ [Fe/H] $\le -0.59$) turned out {\it
  very close to the Solar System value}, $\eps{met}({\rm Pb/Eu}) =
1.53$ \citep{Lodders2009}. This is in agreement with the theoretical
result of \citet{2001ApJ...549..346T} who predicted that the AGB stars
with [Fe/H] $\simeq -1$ made the greatest contribution to the
abundance of s-nuclei of lead in the presolar nebula from which the
Sun originated.


\begin{acknowledgements}
 L.M. is supported by the Swiss National Science Foundation (SCOPES
 project No.~IZ73Z0-128180/1) and the RF President with a grant on Leading Scientific
Schools 3602.2012.2.  A.~F. is, in part, supported by a Clay
 Fellowship administered by the Smithsonian Astrophysical Observatory.
 We used the MARCS model atmosphere library and the NIST and VALD
 databases.
\end{acknowledgements}

\bibliography{ml_pb}
\bibliographystyle{aa}


\end{document}